\documentclass[a4paper,fleqn,usenatbib]{mnras}

\usepackage[T1]{fontenc}
\usepackage{ae,aecompl}

\usepackage{graphicx}	
\usepackage{amsmath}	
\usepackage{amssymb}	
\usepackage{subfigure}
\usepackage{multirow}
\usepackage{xcolor}

\newcommand{\pcc}{\,{\rm cm}^{-3}}
\newcommand{\gcc}{\,{\rm g \, cm}^{-3}}

\newcommand{\kel}{\, {\rm K}}

\newcommand{\msun}{\, {\rm M}_\odot}
\newcommand{\nh}{n_{_{\rm H}}}

\newcommand{\pc}{\, {\rm pc}}

\newcommand{\mh}{m_{\rm H}}

\newcommand{\yr}{\, {\rm yr}}
\newcommand{\myr}{\, {\rm Myr}}

\newcommand{\kms}{\, {\rm km \, s^{-1}}}

\newcommand{\mg}{\, {\rm \mu G}}

\title[Magnetic prestellar cores]{The initial magnetic criticality of prestellar cores}

\author[Priestley et al.]{
Felix D. Priestley$^1$, Charles Yin$^1$ \& James Wurster$^2$
\\
$^1$School of Physics and Astronomy, Cardiff University, Queen's Buildings, The Parade, Cardiff CF24 3AA, UK \\
$^2$SUPA, School of Physics and Astronomy, University of St Andrews, North Haugh, St Andrews, Fife KY16 9SS, UK \\
}

\date{Accepted XXX. Received YYY; in original form ZZZ}

\pubyear{2022}

\begin{document}
\label{firstpage}
\pagerange{\pageref{firstpage}--\pageref{lastpage}}
\maketitle

\begin{abstract}

  Direct observational measurements of the magnetic field strength in prestellar cores {typically find supercritical mass-to-flux ratios}, suggesting that the magnetic field is insufficient to prevent gravitational collapse. These measurements suffer from significant uncertainties; an alternative approach is to utilise the sensitivity of prestellar chemistry to the evolutionary history, {and indirectly} constrain the degree of magnetic support. We combine non-ideal magnetohydrodynamic simulations of prestellar cores with time-dependent chemistry and radiative transfer modelling, producing synthetic observations of the model cores in several commonly-observed molecular lines. We find that molecules {strongly} affected by freeze-out, such as CS and HCN, typically have much lower line intensities {in magnetically subcritical models} compared to supercritical ones, {due to the longer collapse timescales}. {Subcritical models also produce much narrower lines for all species investigated.} Accounting for a range of core properties, ages and viewing angles, we find that supercritical {models} are unable to reproduce the distribution of CS and N$_2$H$^+$ line strengths {and widths} seen in an observational sample, {whereas subcritical models are in good agreement with the available data}. This suggests that despite presently having supercritical mass-to-flux ratios, prestellar cores form {as magnetically subcritical objects.}

\end{abstract}

\section{Introduction}

Stars form in overdense regions of molecular clouds, known as prestellar cores, with typical densities $\nh \gtrsim 10^3 \pcc$ \citep{bergin2007}. While the processes that form these structures are still not fully understood, it is generally assumed that cores are magnetically supercritical, in that their mass-to-flux ratio is above the value where magnetic support can prevent gravitational collapse \citep{mouschovias1976}. Direct observational measurements of the magnetic field strength in molecular clouds \citep{crutcher2012,liu2022,pattle2022} find that at the {relevant densities}, structures are trans- to supercritical, suggesting that {while they play an important role in many areas \citep{wurster2020a,wurster2020b}, magnetic fields are unable to significantly delay (or otherwise regulate) star formation at the scale of prestellar cores.}

Direct measurements of the magnetic field suffer from large uncertainties, due to the faintness of the signals involved and the assumptions required to convert, for example, polarised sub-millimetre emission into a field strength \citep{lyo2021,skalidis2021,pattle2022}. They are also biased towards targets where such a measurement is possible. These are necessarily the brightest objects, and so presumably also the most dense, which suggests that they have already undergone some degree of gravitational contraction. In fact, magnetically-supported models of star formation predict that the central region of a subcritical core becomes supercritical and collapses, due to the imperfect coupling between the magnetic field and the predominantly-neutral gas \citep{fiedler1993}. The fact that cores are observed to have supercritical magnetic field strengths does not exclude the possibility that they form from subcritical initial conditions.\footnote{This is necessarily the case, because the lower-density interstellar medium is found to {be magnetically subcritical}, but the transition to supercriticality {is thought to} occur {well} before the formation of the cores themselves as distinct objects \citep{ching2022}.}

The molecular chemistry of prestellar cores is highly sensitive to their evolutionary history, because the timescales of chemical reactions under these conditions are comparable to, or longer than, those of the other relevant processes \citep{banerji2009}. The significantly longer duration of collapse for subcritical cores compared to supercritical ones \citep{machida2018} alters their chemical makeup, providing an alternative, indirect method of probing the degree of magnetic support. Theoretical studies \citep[e.g.][]{tassis2012,priestley2018,priestley2019,tritsis2021} have found large differences in the molecular abundances of sub- and supercritical models, suggesting their use as a diagnostic of the mass-to-flux ratio. However, uncertainties in both the {measured} abundances and the chemical networks employed, combined with much smaller differences in the observationally-accessible column densities (as opposed to volume densities; \citealt{priestley2019}), have restricted the {utility} of these results.

{An alternative approach is to use the output of chemical models to calculate their predicted line emission properties, which can be directly compared to observations. While this combination of time-dependent chemistry with line radiative transfer has seen much previous use \citep{rawlings1992,keto2015,lin2020}, the underlying physical models in such studies are typically restricted to spherical symmetry \citep[e.g.][]{pagani2013,sipila2018}. Magnetised cores are inherently non-spherical, particularly for subcritical mass-to-flux ratios, requiring multi-dimensional magnetohydrodynamic (MHD) simulations to correctly capture their structure and evolution. These simulations are now frequently coupled with time-dependent chemistry \citep{priestley2019,bovino2021,tritsis2021}, providing all the data required to produce synthetic observations of the model prestellar cores.}

In \citet{yin2021}, we used radiative transfer modelling to convert the three-dimensional chemical structure of cores, the result of combined non-ideal MHD simulations and time-dependent chemistry, into position-position-velocity cubes of the line intensity for a number of commonly-observed molecular lines. We found that the ratio of the CS $J=2-1$ and N$_2$H$^+$ $J=1-0$ rotational transitions could robustly distinguish between our subcritical and supercritical models. {Application of} this diagnostic to the well-studied core L1498 suggested that it formed from subcritical initial conditions, despite a direct measurement of the magnetic field strength finding a supercritical mass-to-flux ratio \citep{kirk2006}. However, our analysis was limited to a single model core (with two values of the initial mass-to-flux ratio); it is possible that changing the initial density, for example, could reconcile supercritical models with the molecular line data. In this paper, we expand our modelling to include cores of different masses, densities, ages and orientations, confirming the {basic} validity of the CS/N$_2$H$^+$ diagnostic proposed in \citet{yin2021}. A comparison of our model results with an extended observational dataset suggests that most, if not all, cores form {as magnetically subcritical objects}, in contrast to the conclusions drawn from direct field strength measurements.

\section{Method}

\begin{table*}
  \centering
  \caption{Initial core radius, mass, density of hydrogen nuclei, magnetic field strength {and mass-to-flux ratio} for each of our models, the duration of the simulation, {and the initial free-fall and ambipolar diffusion timescales \citep{hartquist1989,banerji2009}}.}
  \begin{tabular}{ccccccccc}
    \hline
    Model & $R$/pc & $M$/$\msun$ & $\nh$/$\pcc$ & $B_z$/$\mg$ & $\mu$/$\mu_{\rm crit}$ & $t_{\rm end}$/Myr & $t_{\rm ff}$/Myr & $t_{\rm AD}$/Myr \\
    \hline
    D3L6-SUP & $0.59$ & $50$ & $1.7 \times 10^3$ & $4.2$ & $4.4$ & $1.06$ & $1.82$ & $2.91$ \\
    D3L6-SUB & $0.59$ & $50$ & $1.7 \times 10^3$ & $42$ & $0.44$ & $1.59$ & $1.82$ & $2.91$ \\
    D4L1-SUP & $0.13$ & $5$ & $1.6 \times 10^4$ & $9.1$ & $4.4$ & $0.306$ & $0.592$ & $0.758$ \\
    D4L1-SUB & $0.13$ & $5$ & $1.6 \times 10^4$ & $91$ & $0.44$ & $0.965$ & $0.592$ & $0.758$ \\
    D4L3-SUP & $0.28$ & $50$ & $1.6 \times 10^4$ & $19.6$ & $4.4$ & $0.329$ & $0.592$ & $0.758$ \\
    D4L3-SUB & $0.28$ & $50$ & $1.6 \times 10^4$ & $196$ & $0.44$ & $0.428$ & $0.592$ & $0.758$ \\
     \hline
  \end{tabular}
  \label{tab:mhdprop}
\end{table*}

\begin{table}
  \centering
  \caption{{Initial gas-phase elemental abundances, relative to hydrogen nuclei, used in the chemical modelling.}}
  \begin{tabular}{cccc}
    \hline
    Element & Abundance & Element & Abundance \\
    \hline
    C & $7.3 \times 10^{-5}$ & S & $8.0 \times 10^{-6}$ \\
    N & $2.1 \times 10^{-5}$ & Si & $8.0 \times 10^{-7}$ \\
    O & $1.8 \times 10^{-4}$ & Mg & $7.0 \times 10^{-7}$ \\
    \hline
  \end{tabular}
  \label{tab:abun}
\end{table}

\begin{table}
  \centering
  \caption{{Molecules for which we perform radiative transfer modelling, the collisional partners used, and references for the collisional rates.}}
  \begin{tabular}{ccc}
    \hline
    Molecules & Partner & Reference \\
    \hline
    CS & H$_2$ & \citet{lique2006}\\
    N$_2$H$^+$ & H$_2$ & \citet{flower1999} \\
    HCN & H$_2$ & \citet{dumouchel2010} \\
    HCO$^+$ & H$_2$ & \citet{flower1999} \\
    \hline
  \end{tabular}
  \label{tab:moldata}
\end{table}

\begin{figure*}
  \centering
  \includegraphics[width=\columnwidth]{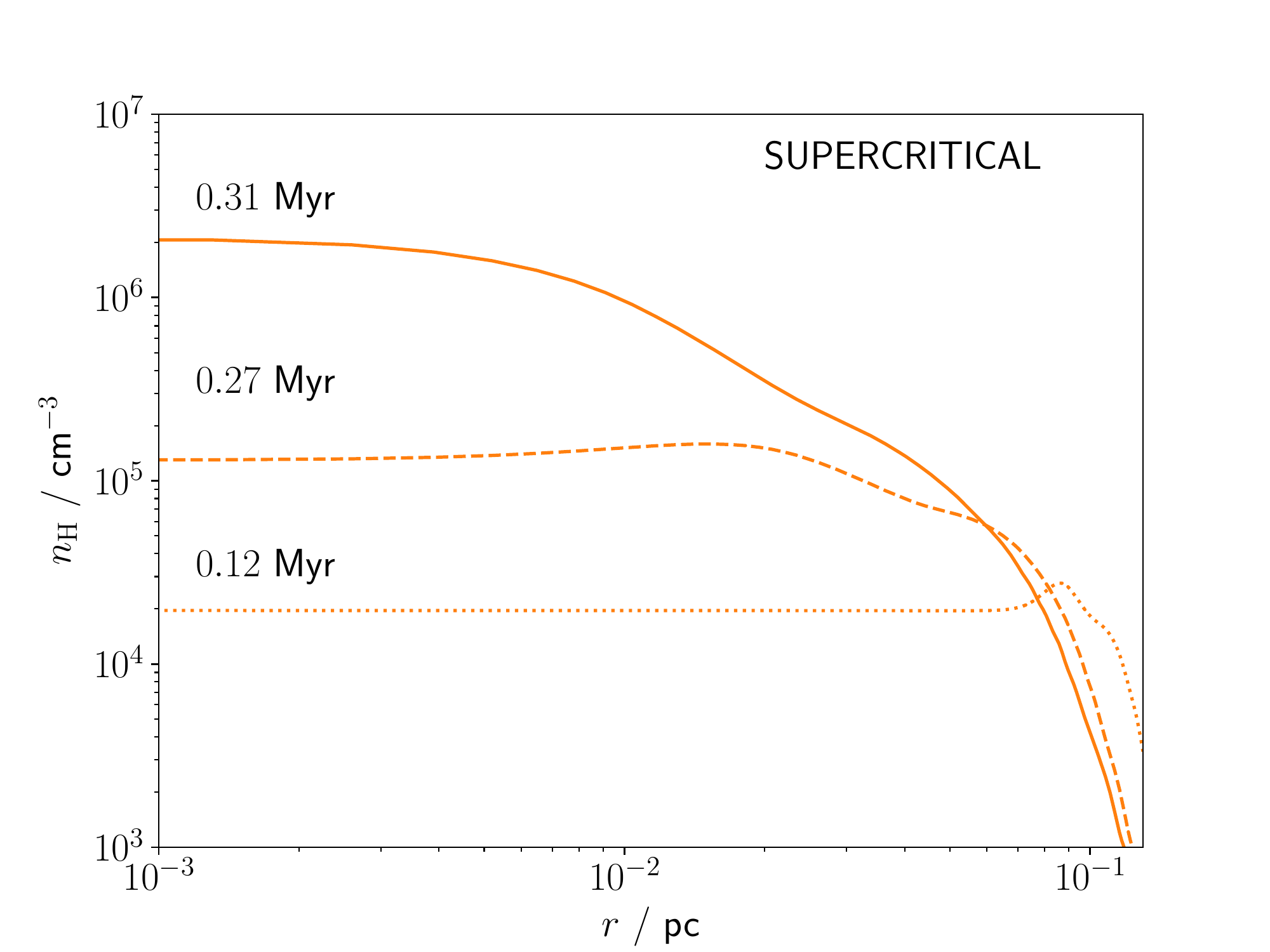}\quad
  \includegraphics[width=\columnwidth]{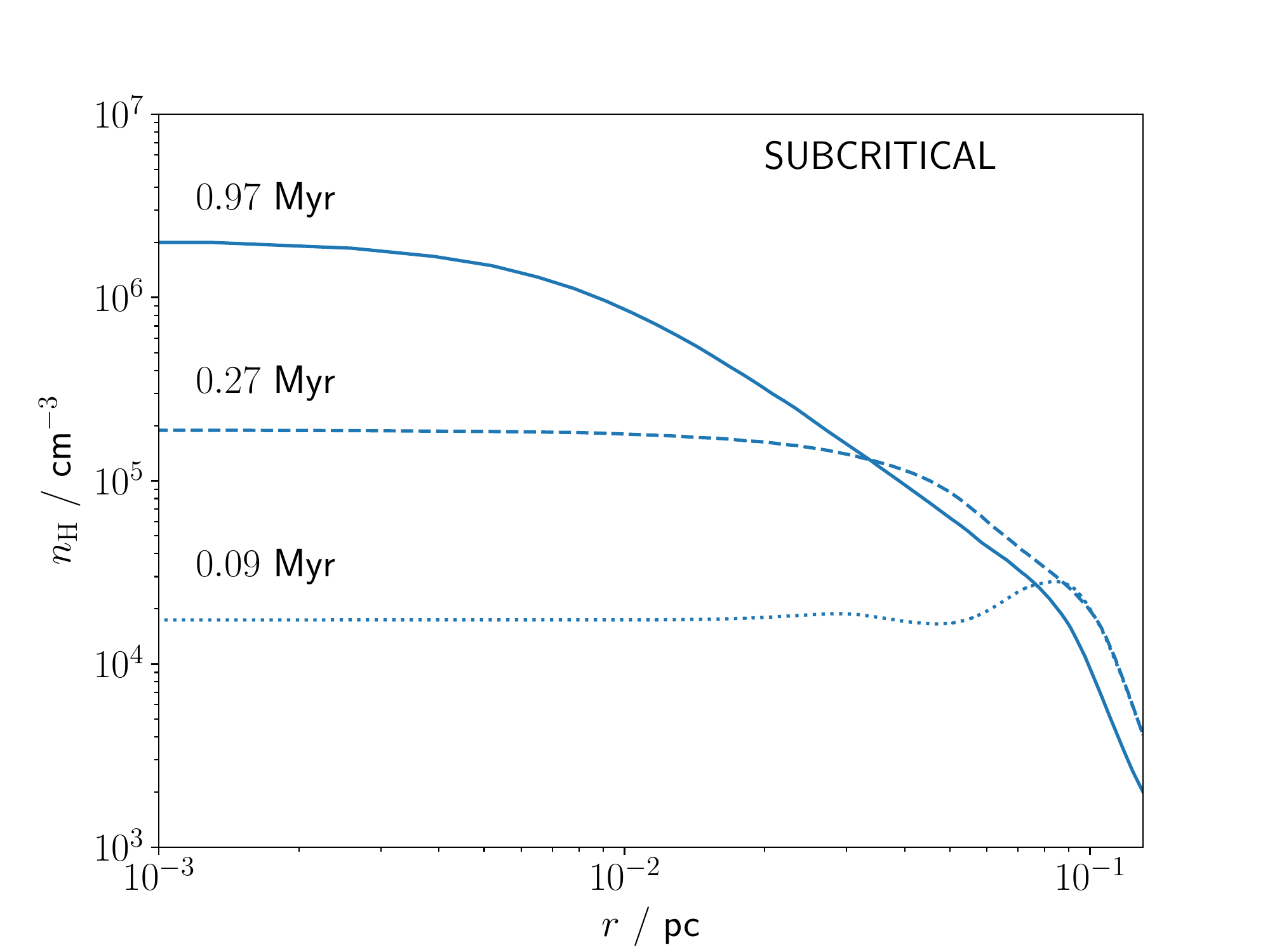}\quad
  \caption{{Midplane volume density profiles for the D4L1-SUP (left) and D4L1-SUB (right) models, at the point where the models have reached central densities of approximately $2 \times 10^4 \pcc$ (dotted lines), $2 \times 10^5 \pcc$ (dashed lines), and $2 \times 10^6 \pcc$ (solid lines). Times are indicated above the profiles.}}
  \label{fig:dens}
\end{figure*}

\begin{figure*}
  \centering
  \includegraphics[width=\columnwidth]{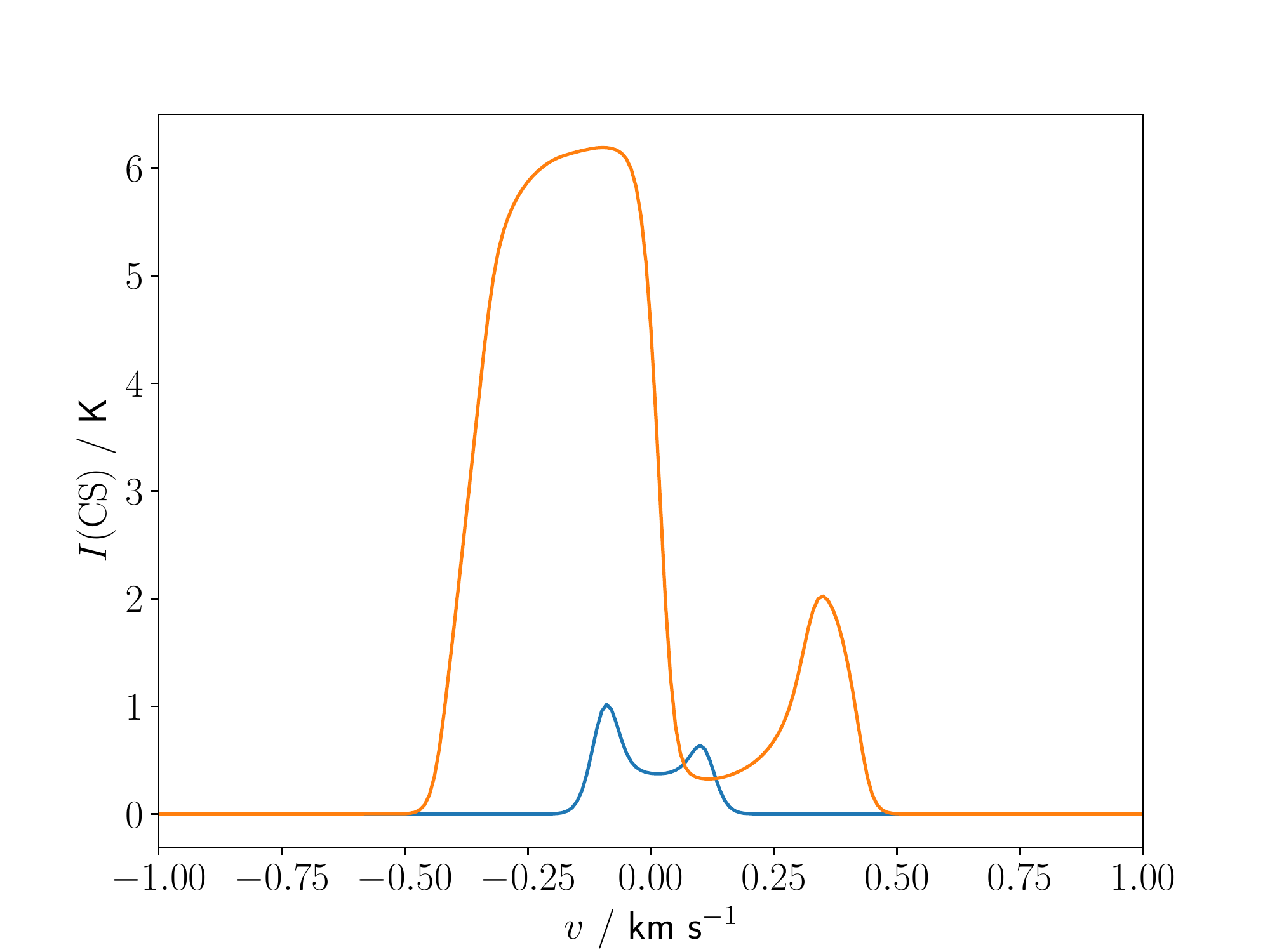}\quad
  \includegraphics[width=\columnwidth]{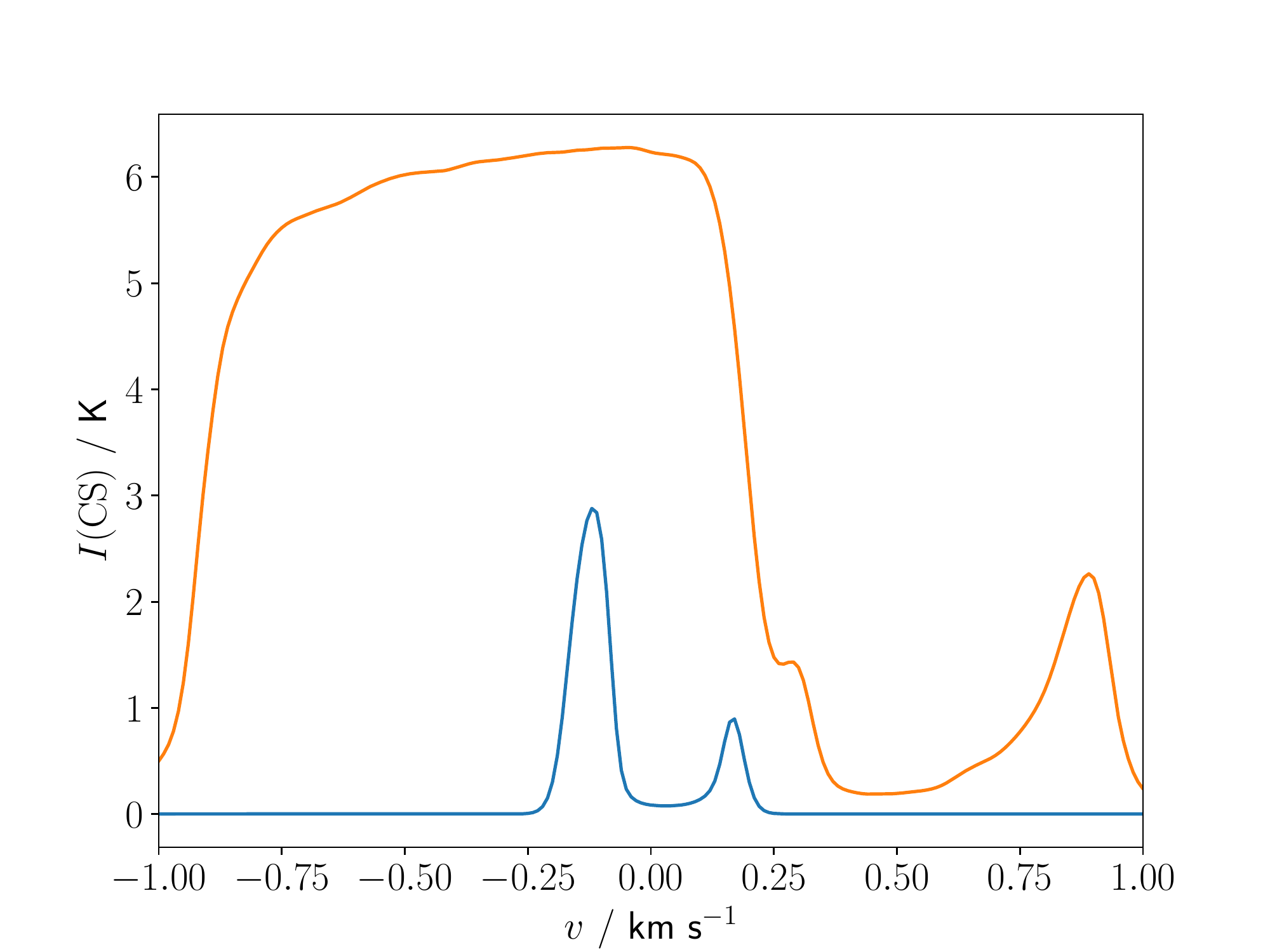}\quad
  \caption{{CS $J=2-1$ line profiles, viewed perpendicular to the initial magnetic field direction at $t_{\rm end}$. {\it Left:} the D4L1-SUP model at $0.30 \myr$ (orange line), and the D4L1-SUB model at $0.97 \myr$ (blue line). {\it Right:} the D3L6-SUP model at $1.1 \myr$ (orange line), and the D3L6-SUB model at $1.6 \myr$ (blue line).}}
  \label{fig:cs}
\end{figure*}

\begin{figure*}
  \centering
  \includegraphics[width=\columnwidth]{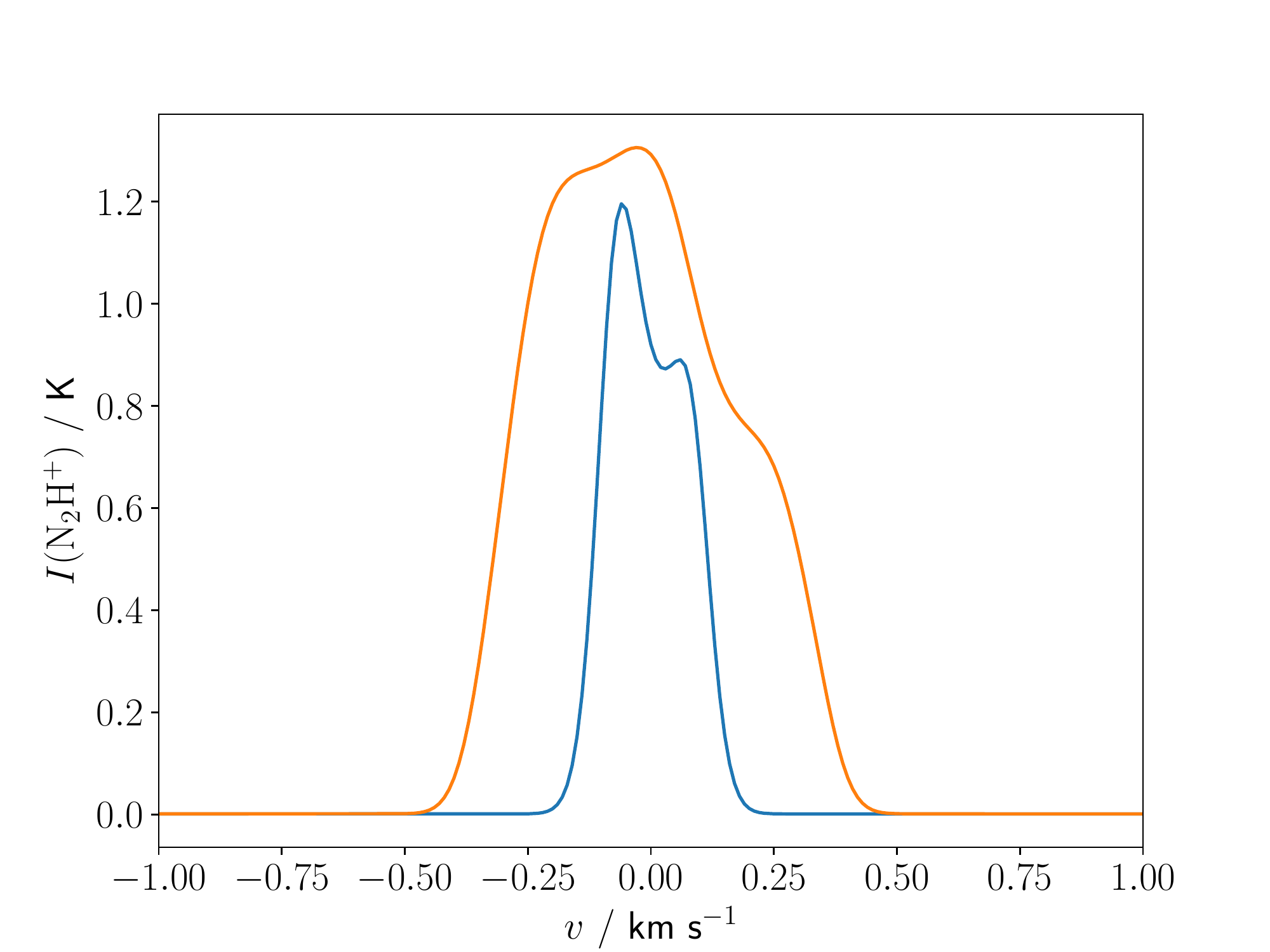}\quad
  \includegraphics[width=\columnwidth]{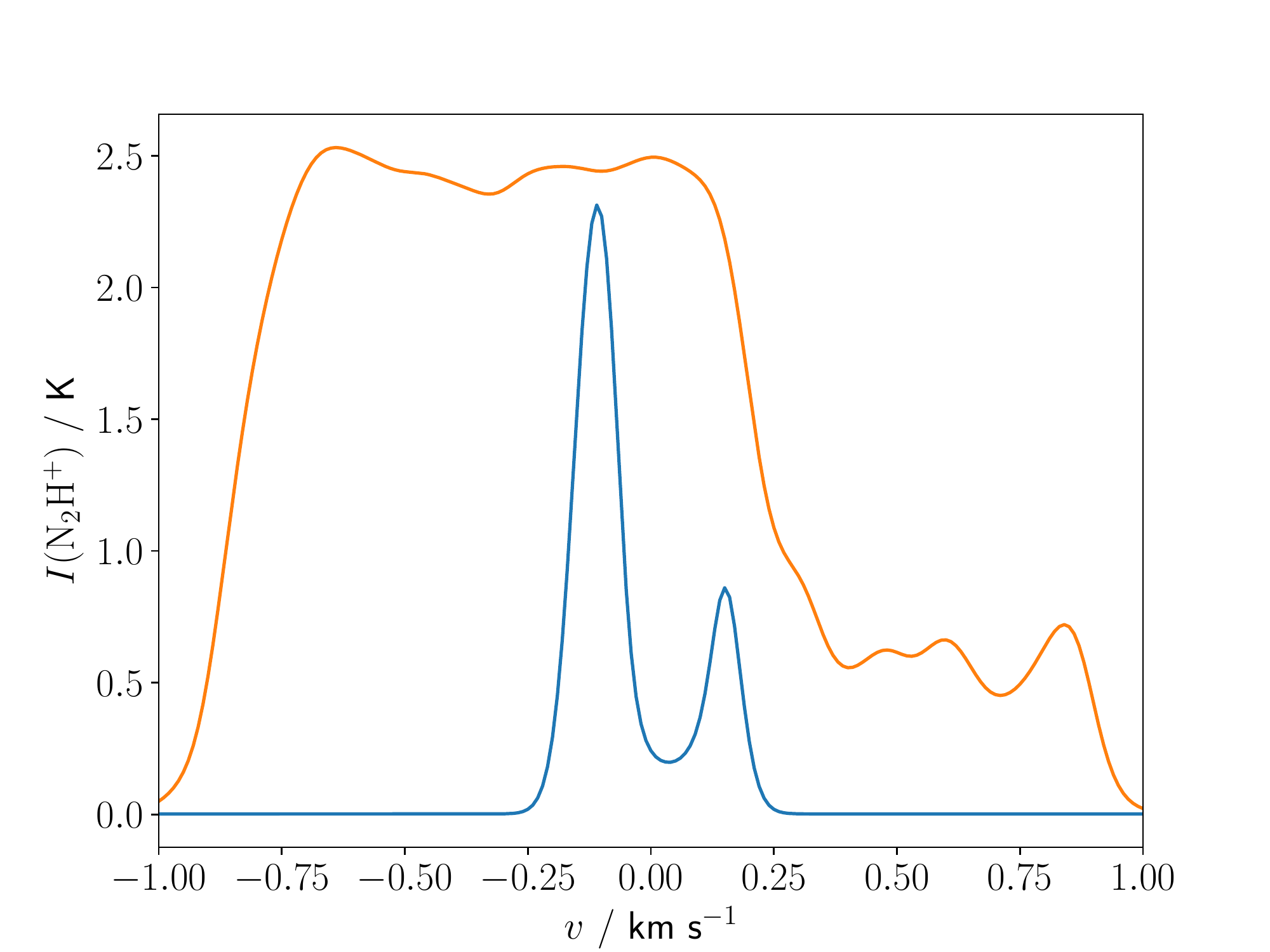}\quad
  \caption{{N$_2$H$^+$ $J=1-0$ line profiles, viewed perpendicular to the initial magnetic field direction at $t_{\rm end}$. {\it Left:} the D4L1-SUP model at $0.30 \myr$ (orange line), and the D4L1-SUB model at $0.97 \myr$ (blue line). {\it Right:} the D3L6-SUP model at $1.1 \myr$ (orange line), and the D3L6-SUB model at $1.6 \myr$ (blue line).}}
  \label{fig:n2h}
\end{figure*}

We simulate the non-ideal MHD evolution of prestellar cores using {\sc phantom}, a smoothed-particle (magneto)hydrodynamics (SPH) code \citep{price2018}. Our initial conditions are static, uniform-density spheres with a uniform magnetic field in the $z$-direction. Non-ideal coefficients are calculated using the {\sc nicil} library {(version $1.2.6$; \citealt{wurster2016}), with an ion mass of $24.3 \, \mh$ and} the ion density replaced by the expression
\begin{equation}
  n_i/\nh = 10^{-7} \left(\frac{\nh}{10^3 \pcc}\right)^{-0.6},
\end{equation}
which approximates the output of time-dependent chemical networks well for the range of densities investigated \citep{tassis2012b,priestley2019}. The sound speed in the cores is set to $c_s = 0.2 \kms$, corresponding to molecular gas at $\sim 10 \kel$, and the cores are surrounded by a background medium with a density $30$ times lower, and sound speed $30^{1/2}$ times higher, to provide a confining pressure. {We use an isothermal equation of state, and assume the core material is isothermal at $10 \kel$ throughout the rest of the paper.} We run the models with $200 \, 000$ particles\footnote{{We investigate the effect of using higher resolutions in Appendix \ref{sec:res}, finding that it has a negligible impact on our results.}}, and terminate the evolution when the maximum density reaches $4.7 \times 10^{-18} \gcc$ (i.e. $\nh = 2 \times 10^6 \pcc$). Model parameters are listed in Table \ref{tab:mhdprop}.

We consider three combinations of core mass and radius, covering masses of $5$ and $50 \msun$ and initial densities in the range $\sim 10^3-10^4 \pcc$. For each of these setups, we run the model once with a subcritical and once with a supercritical magnetic field strength, giving a total of six MHD simulations. We consider mass-to-flux ratios of $4.4$ ($0.44$) times the \citet{mouschovias1976} critical value for the supercritical (subcritical) models.\footnote{These values were incorrectly stated to be $5.0$ and $0.5$ in \citet{yin2021}.} The D4L1 pair of models are those studied previously in \citet{yin2021}; we here investigate whether those results generalise to cores with different initial conditions. Model parameters are listed in Table \ref{tab:mhdprop}.

For each model, we post-process the chemical evolution of $10 \, 000$ randomly-selected particles using {\sc uclchem} \citep{holdship2017}, with the {\sc umist12} reaction network \citep{mcelroy2013}, high-metal elemental abundances from \citet{lee1998} {(listed in Table \ref{tab:abun})}, a constant gas/grain temperature of $10 \kel$, a cosmic ray ionization rate of $1.3 \times 10^{-17} \, {\rm s^{-1}}$, and no external radiation field, as cores are typically well-shielded by their parent molecular clouds. {We assume hydrogen is initially in molecular form, while all other elements are atomic, and} allow the chemistry to evolve at the initial core density for $10^4 \yr$ before beginning the collapse.

We use the three-dimensional chemical structure of the cores as input for {\sc lime} \citep{brinch2010}, a non-local thermodynamic equilibrium line radiative transfer code, with molecular properties taken from the {\sc lamda} database\footnote{{Some of the molecular lines we consider have hyperfine structure, which is not accounted for in the default {\sc lamda} data files. In Appendix \ref{sec:hfs}, we find that the inclusion of hyperfine structure is unlikely to substantially alter our conclusions.}} \citep{schoier2005} and dust properties from \citet{ossenkopf1994}. We use $10 \, 000$ sample points, and assign each one the properties of the nearest post-processed SPH particle. {Molecular species considered, their collisional partners, and references for the collisional data are given in Table \ref{tab:moldata}.} We perform radiative transfer modelling at $t_{\rm end}$ and $t_{\rm end}/2$, and for viewing angles of $0^\circ$ and $90^\circ$ to the initial magnetic field direction, and extract line profiles by averaging the intensity within $0.05 \pc$ of the centre. We thus have four line profiles for each model in Table \ref{tab:mhdprop}, eight for each core mass/size combination, and a total of $24$ for each transition considered, covering a range of both physical core properties and observational circumstances.

\section{Results}

\begin{figure*}
  \centering
  \includegraphics[width=0.4\textwidth]{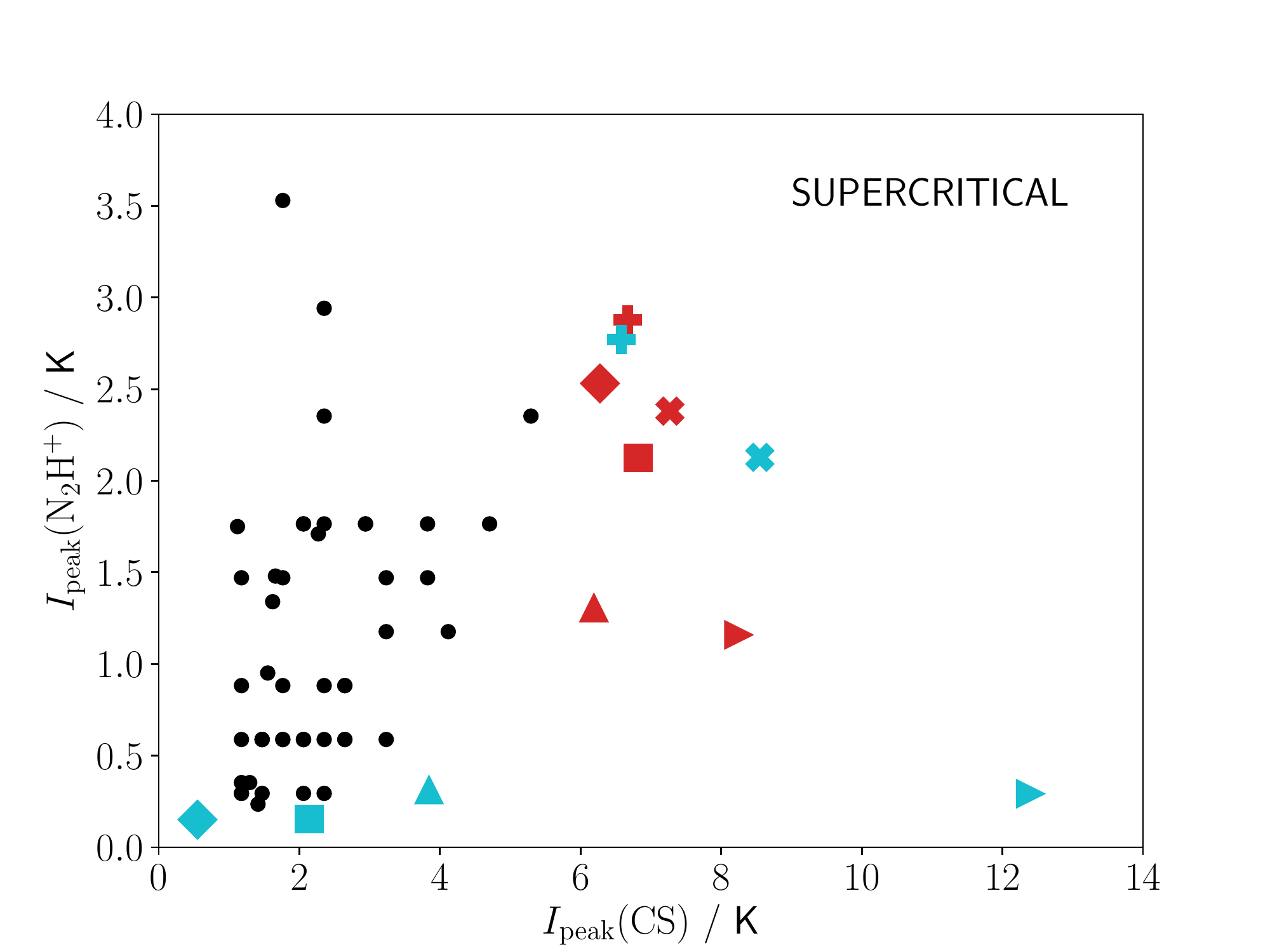}\quad
  \includegraphics[width=0.4\textwidth]{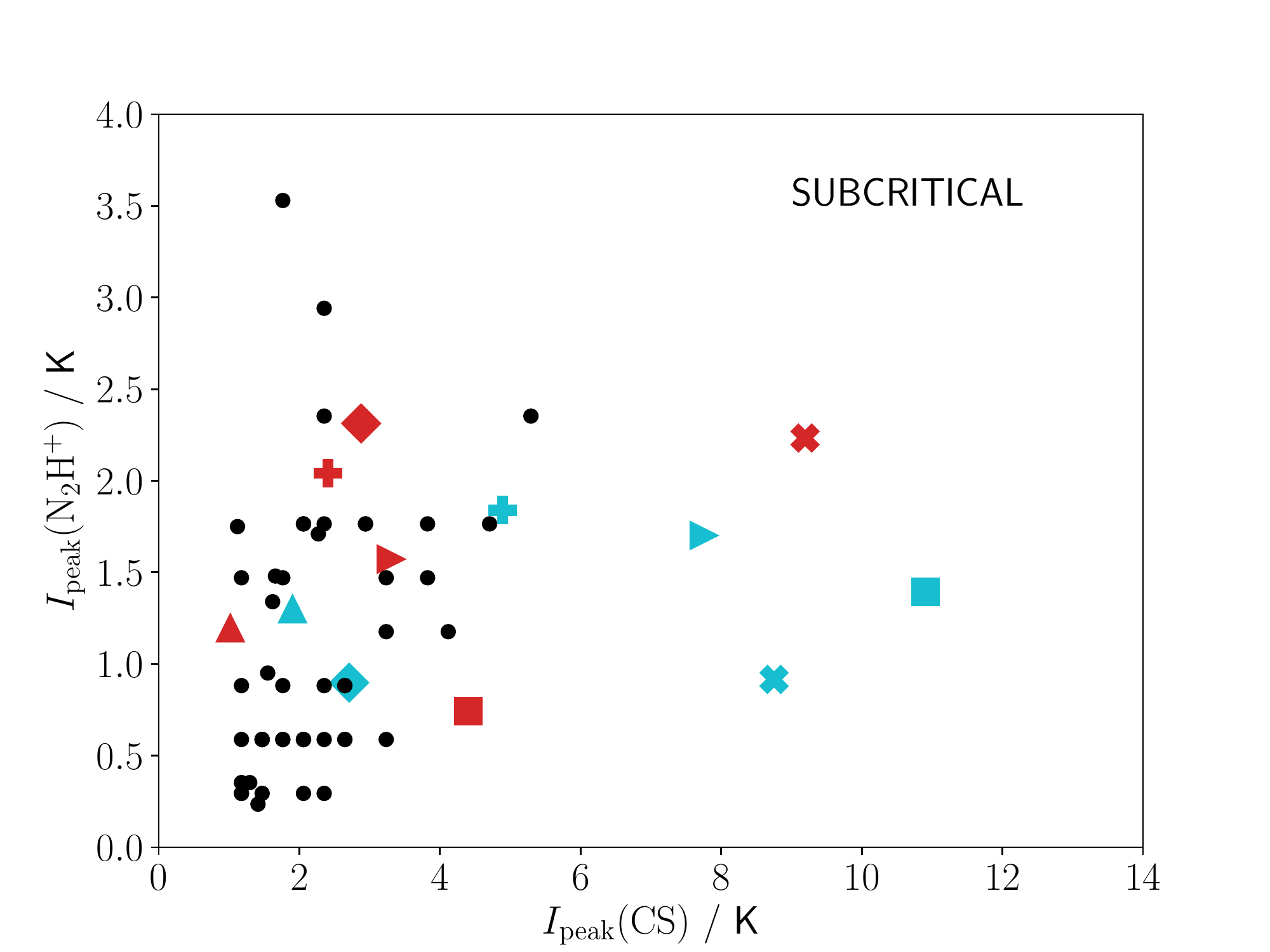}\quad
  \includegraphics[width=0.11\textwidth]{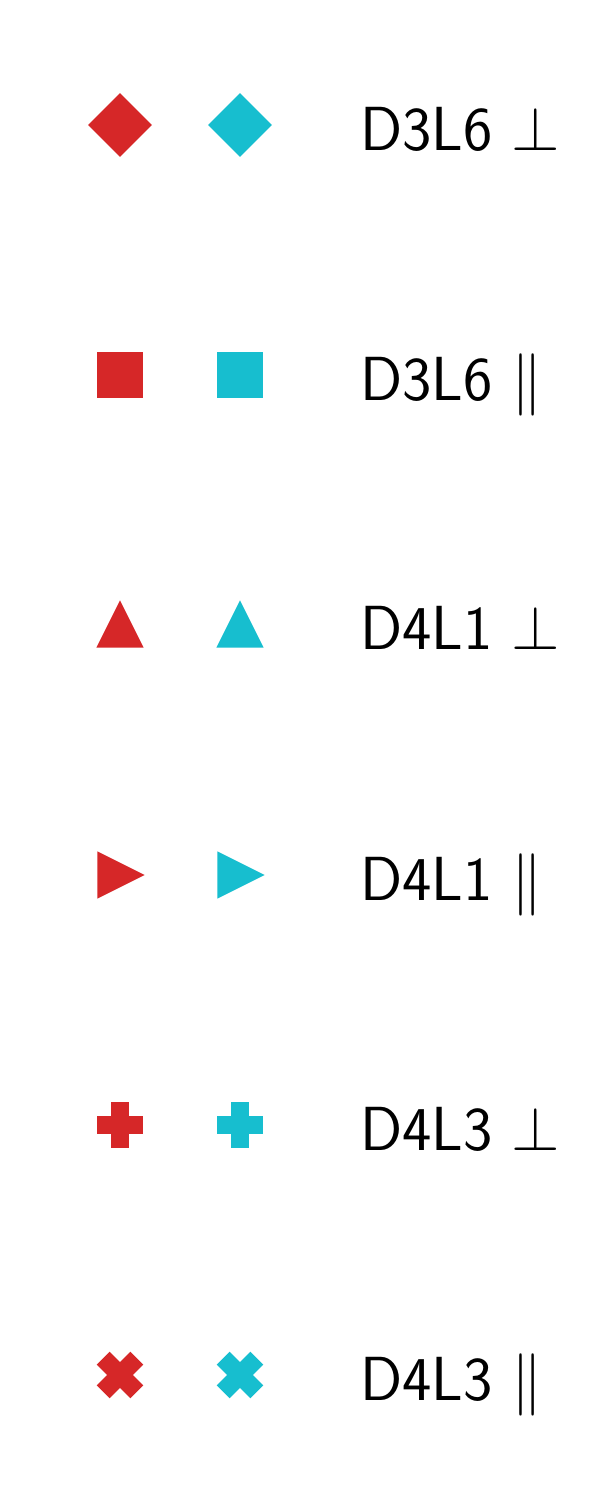}\quad
  \caption{{Peak intensity of the CS $J=2-1$ line versus the N$_2$H$^+$ $J=1-0$ line for supercritical (left panel) and subcritical (right panel) models. Red symbols are models at $t_{\rm end}$, cyan symbols those at $t_{\rm end}/2$. Symbol shapes represent the physical model, and whether the viewing angle is parallel or perpendicular to the initial magnetic field direction. Observed peak line intensities from \citet{lee1999} and \citet{tafalla2002} are shown as black circles.}}
  \label{fig:csflux}
\end{figure*}

\begin{figure*}
  \centering
  \includegraphics[width=0.4\textwidth]{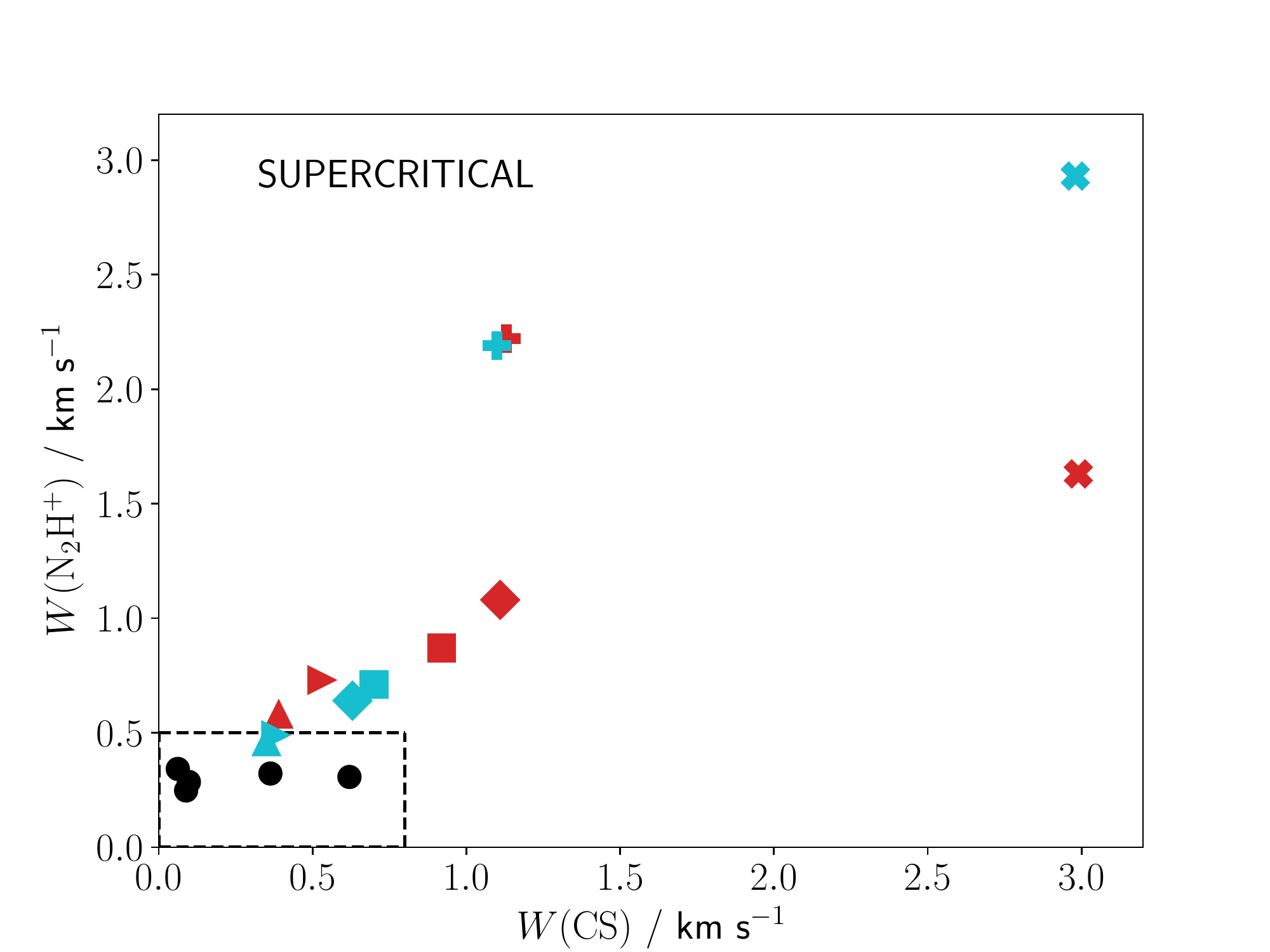}\quad
  \includegraphics[width=0.4\textwidth]{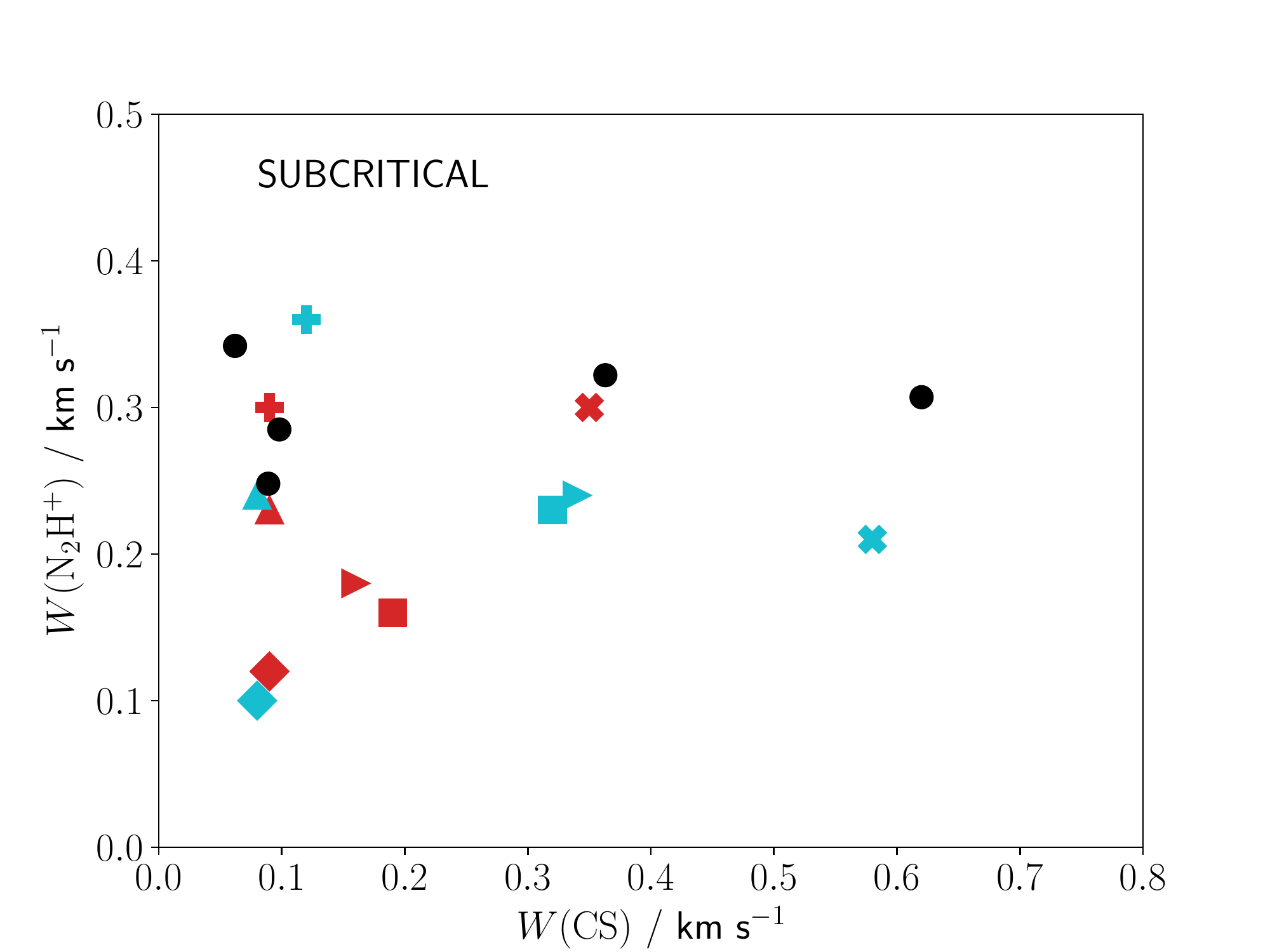}\quad
  \includegraphics[width=0.11\textwidth]{legend}\quad
  \caption{{Full-width at half-maximum of the CS $J=2-1$ line versus the N$_2$H$^+$ $J=1-0$ line for supercritical (left panel) and subcritical (right panel) models. Red symbols are models at $t_{\rm end}$, cyan symbols those at $t_{\rm end}/2$. Symbol shapes represent the physical model, and whether the viewing angle is parallel or perpendicular to the initial magnetic field direction. Observed line widths from \citet{tafalla2002} are shown as black circles. The dashed black rectangle in the left panel indicates the size of the region covered by the right panel.}}
  \label{fig:cswidth}
\end{figure*}

\begin{figure*}
  \centering
  \includegraphics[width=0.4\textwidth]{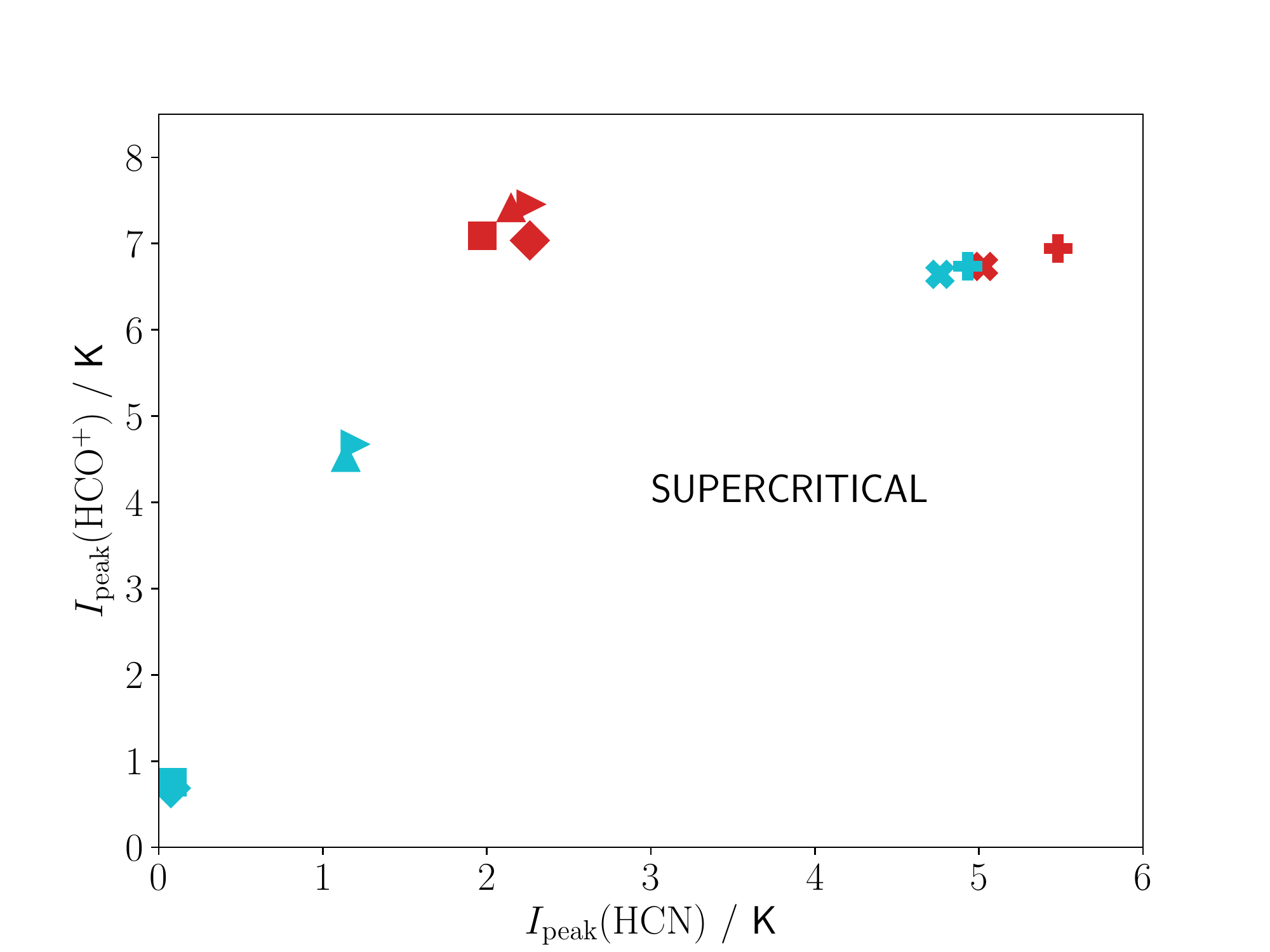}\quad
  \includegraphics[width=0.4\textwidth]{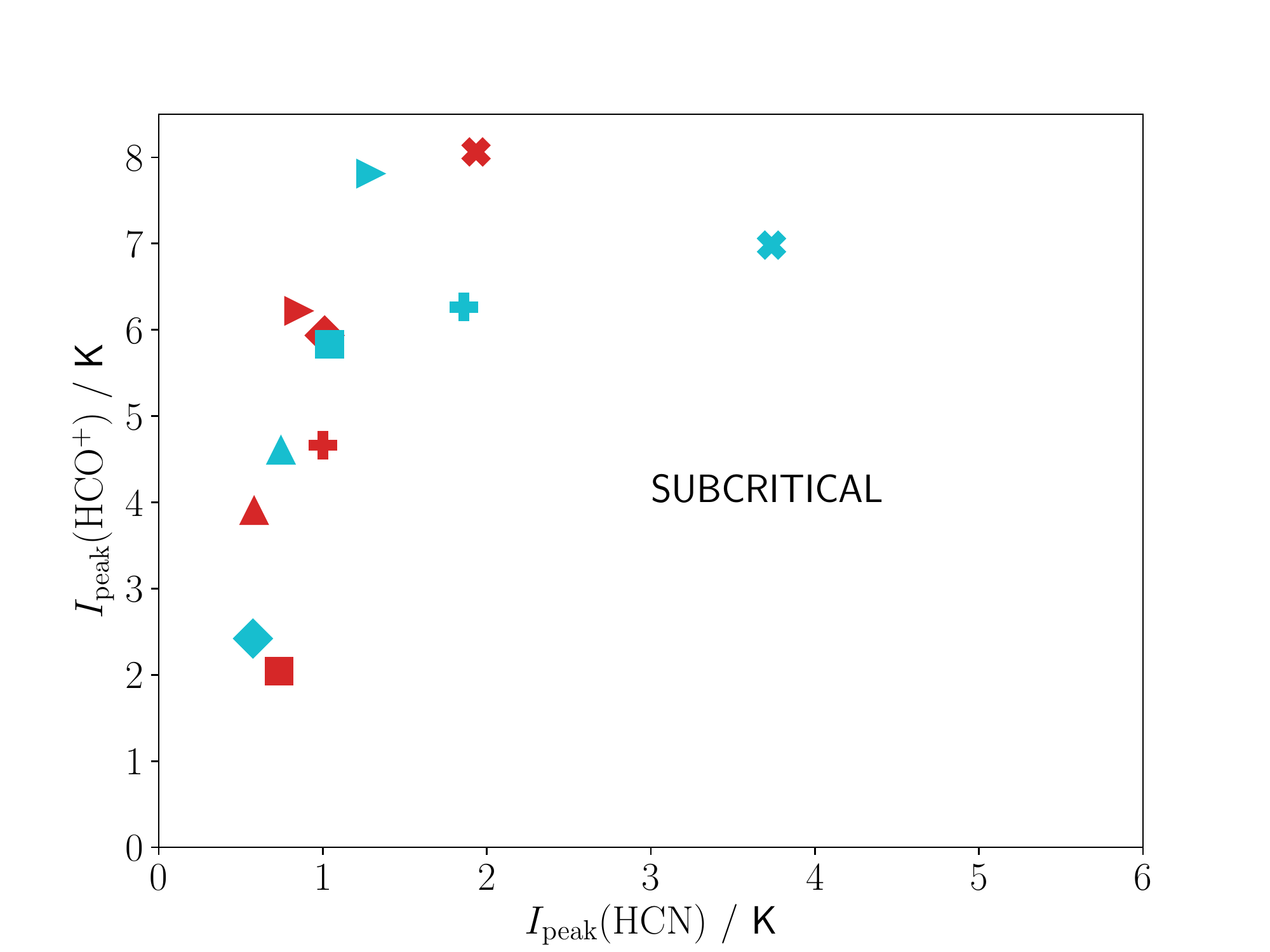}\quad
  \includegraphics[width=0.11\textwidth]{legend}\quad
  \caption{{Peak intensity of the HCN $J=1-0$ line versus the HCO$^+$ $J=1-0$ line for supercritical (left panel) and subcritical (right panel) models. Red symbols are models at $t_{\rm end}$, cyan symbols those at $t_{\rm end}/2$. Symbol shapes represent the physical model, and whether the viewing angle is parallel or perpendicular to the initial magnetic field direction.}}
  \label{fig:hcnflux}
\end{figure*}

\begin{figure*}
  \centering
  \includegraphics[width=0.4\textwidth]{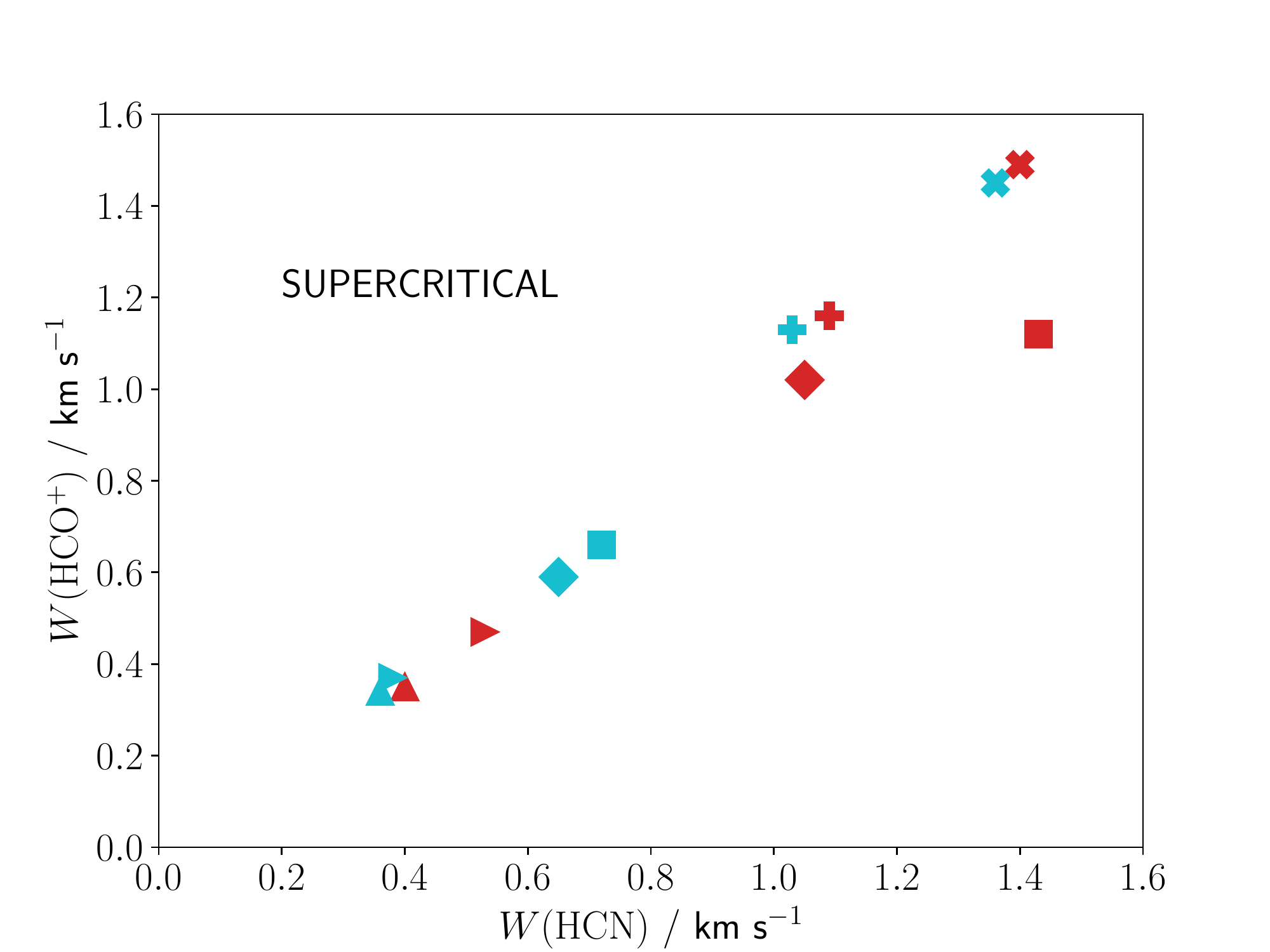}\quad
  \includegraphics[width=0.4\textwidth]{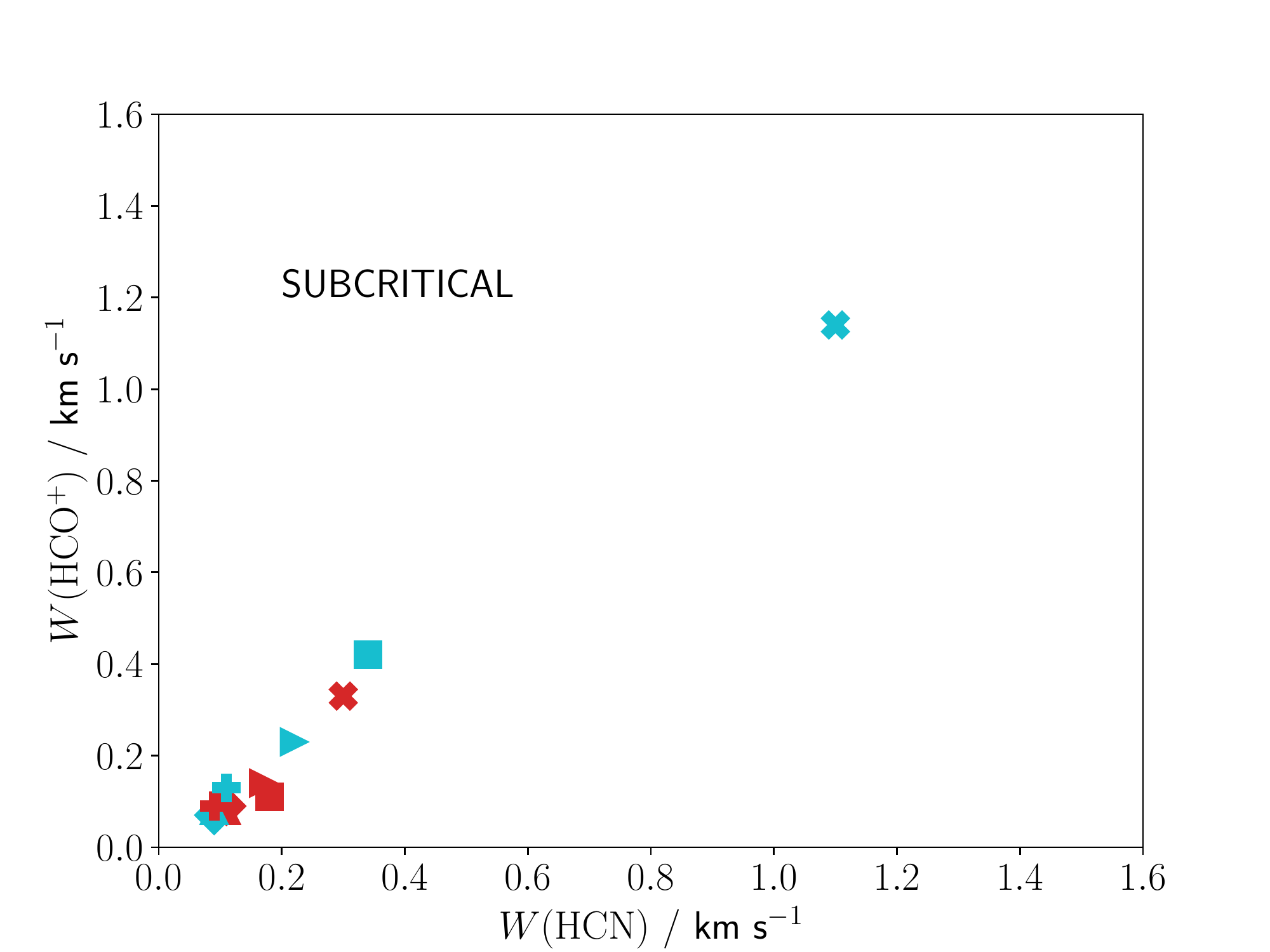}\quad
  \includegraphics[width=0.11\textwidth]{legend}\quad
  \caption{{Full-width at half-maximum of the HCN $J=1-0$ line versus the HCO$^+$ $J=1-0$ line for supercritical (left panel) and subcritical (right panel) models. Red symbols are models at $t_{\rm end}$, cyan symbols those at $t_{\rm end}/2$. Symbol shapes represent the physical model, and whether the viewing angle is parallel or perpendicular to the initial magnetic field direction.}}
  \label{fig:hcnwidth}
\end{figure*}

{Figure \ref{fig:dens} shows the evolution of the volume density profile in the $x-y$ plane (i.e. perpendicular to the original magnetic field direction) for the D4L1 models. Initially, the subcritical and supercritical models evolve on similar timescales, reaching a central density of $2 \times 10^5 \pcc$ in $\sim 0.3 \myr$ in both cases. Beyond this point, their behaviour diverges. The supercritical model continues to collapse dynamically, reaching the final density of $2 \times 10^6 \pcc$ in another $0.04 \myr$. The subcritical model is stabilised by magnetic support, and remains in this configuration with little change for over $0.6 \myr$, at which point enough flux has been removed from the centre of the core for it to collapse. The subcritical core thus spends an order of magnitude more time at the high gas densities where freeze-out onto grain surfaces is effective.}

Figure \ref{fig:cs} shows CS $J=2-1$ line profiles from the D4L1 and D3L6 models {at $t_{\rm end}$, when the models reach a central density of $\nh = 2 \times 10^6 \pcc$}. The longer duration of collapse for magnetically subcritical models tends to increase the amount of freeze-out for susceptible molecules such as CS, resulting in significantly {lower} peak line intensities. Supercritical models also have much broader line profiles, due to the {relatively} unimpeded collapse and subsequent supersonic infall. Species which are {less affected} by depletion, such as N$_2$H$^+$, are also less affected by the magnetic criticality of the core. The peak N$_2$H$^+$ line intensities {(neglecting the line's hyperfine structure; see Appendix \ref{sec:hfs})} in Figure \ref{fig:n2h} are comparable between sub- and supercritical models, although supercritical line profiles are still broader, particularly for the lower-density D3L6 model.

We noted in \citet{yin2021} that this different chemical behaviour could be used to observationally distinguish cores with significant magnetic support from those dominated by self-gravity. Figure \ref{fig:csflux} shows the peak intensity of the CS $J=2-1$ line versus the N$_2$H$^+$ $J=1-0$ line for all 24 core/field strength/orientation/age combinations. {The regions of parameter space occupied by supercritical and subcritical models overlap, so the intensity ratios we proposed as diagnostics in \citet{yin2021} (corresponding to straight lines in the $I_{\rm peak}({\rm CS})-I_{\rm peak}({\rm N_2H^+})$ plane) cannot unambiguously identify the magnetic criticality of an individual core. {Moreover, the same physical model can appear in a completely different region of parameter space if the epoch of observation or viewing angle are different, with no obvious systematic dependence of the line intensities on either property.} However, there is a clear preference for supercritical models to have higher CS intensities, with only three combinations of age, orientation, and initial core properties producing $I_{\rm peak}({\rm CS}) < 5 \kel$ (and all three are at $t_{\rm end}/2$). By contrast, the majority of subcritical models have peak intensities below this value. Line intensities from the samples of prestellar cores presented in \citet{lee1999} and \citet{tafalla2002}\footnote{See Appendix \ref{sec:obs} for details on the observational sample. We convert antenna temperatures in \citet{lee1999} to beam temperatures with an assumed efficiency of $0.17$ taken from that paper.} fall in a region of parameter space occupied almost exclusively by subcritical models, but have little, if any, overlap with supercritical ones.}

{As noted above, supercritical models tend to produce lines which are broader in addition to being stronger. Figure \ref{fig:cswidth} shows CS and N$_2$H$^+$ line full-widths at half maximum (FWHMs). Supercritical and subcritical models occupy almost entirely separate regions of the FWHM-FWHM plots (note the different axis scales), with supercritical model FWHMs extending up to several $\kms$, whereas almost all subcritical models have subsonic ($\lesssim 0.4 \kms$) line widths. The five cores from \citet{tafalla2002} correspond extremely well to the subcritical models, and poorly to the supercritical ones. While CS line FWHMs for the cores in \citet{lee1999} are not available to us, of the 69 objects with N$_2$H$^+$ line FWHMs provided, only 11 ($15 \%$) have a width greater than $0.4 \kms$. A single observed core (BS-1) has a FWHM above $0.6 \kms$, whereas only two supercritical models fall below this boundary. Molecular line properties from supercritical models appear to be very different to those observed in real prestellar cores.}

{While we have so far focused on the CS and N$_2$H$^+$ transitions identified as promising diagnostics in \citet{yin2021}, similar effects can also be seen for other commonly-observed lines. Figures \ref{fig:hcnflux} and \ref{fig:hcnwidth} show the peak intensity and FWHM data for the HCN and HCO$^+$ $J=1-0$ transitions. HCN, like CS, is strongly affected by freeze-out, and so subcritical models tend to display lower peak intensities than supercritical ones. HCO$^+$ is less affected, and like N$_2$H$^+$ therefore covers a similar range of peak intensity regardless of the initial mass-to-flux ratio. {As in Figure \ref{fig:csflux}, line properties for the same underlying physical model can change significantly if the epoch of observation or viewing angle are different, although for these lines, the supercritical models (closer to spherical symmetry than the subcritical ones) are only weakly affected by the orientation of the core.} The line widths are similar to those in Figure \ref{fig:cswidth} - supercritical models almost all have FWHMs greater than $0.5 \kms$, while subcritical models are almost all below this value. This suggests a similar test to those in Figures \ref{fig:csflux} and \ref{fig:cswidth} can be performed using alternative pairs of lines, given the availability of observational data. With modern surveys often covering hundreds of cores in transitions from tens of species \citep[e.g.][]{pety2017,kauffmann2017,barnes2020}, it should be possible to conduct multiple such tests with large samples of objects, mitigating the uncertainties in the chemical and radiative properties of any particular molecule.}

\section{Discussion}
\label{sec:discussion}

\begin{figure*}
  \centering
  \includegraphics[width=\columnwidth]{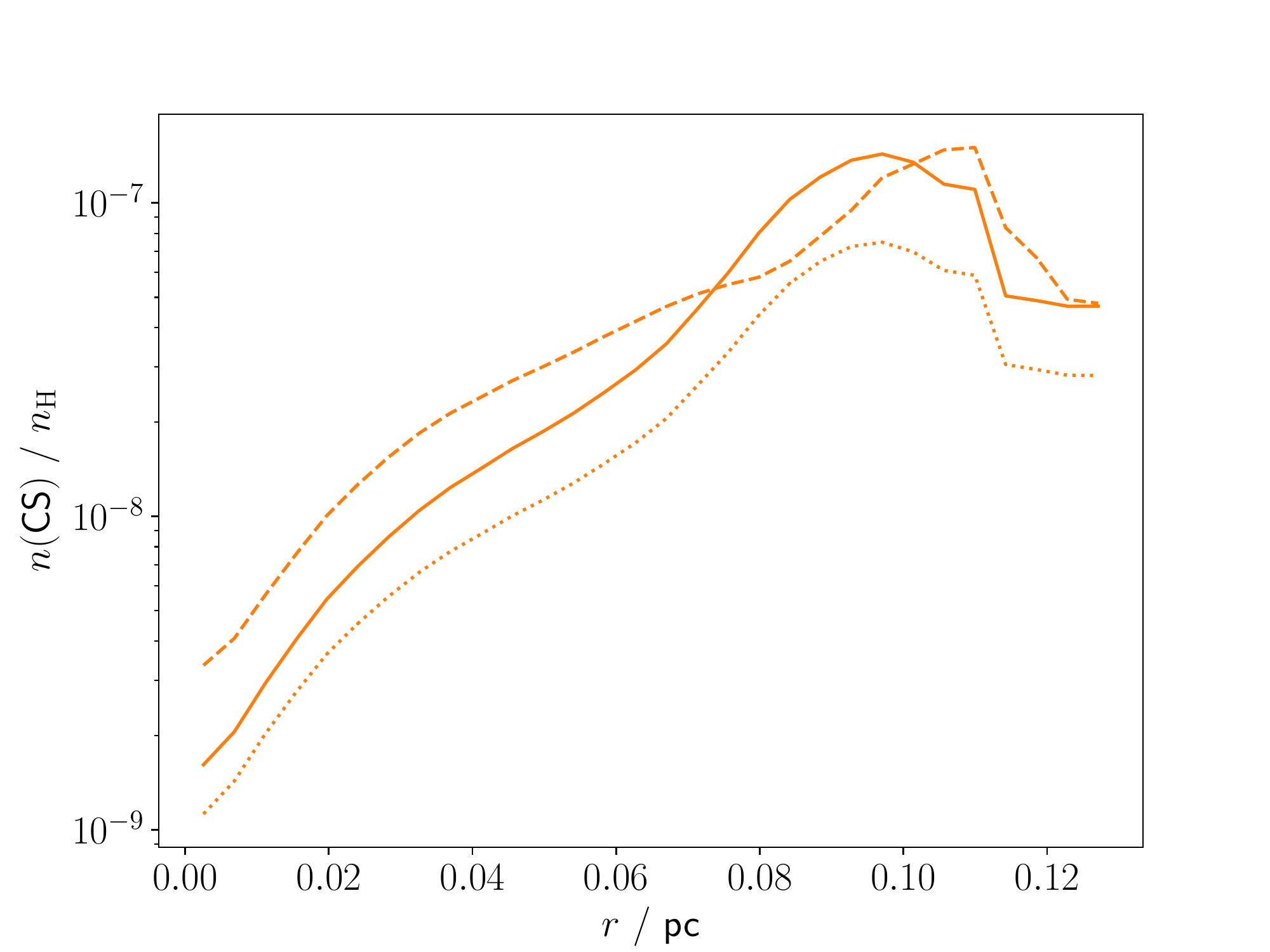}\quad
  \includegraphics[width=\columnwidth]{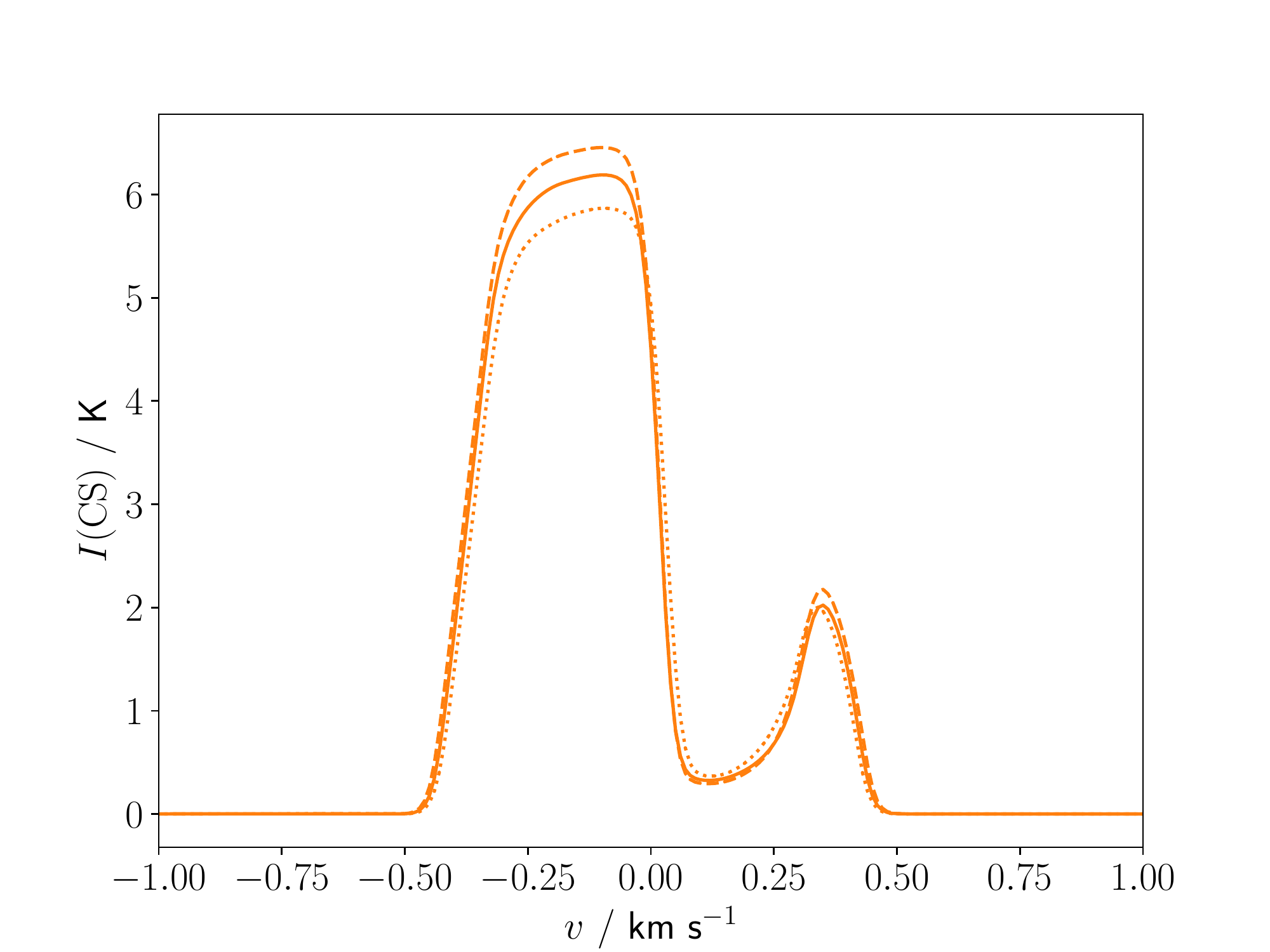}\quad
  \caption{{CS midplane abundance profiles (left) and $J=2-1$ line profiles (right)} from the D4L1-SUP model, viewed {perpendicular to the initial magnetic field direction {at $0.30 \myr$}}, with $10^4$ (solid lines)/$10^5 \yr$ (dashed lines) of preliminary evolution at the initial density, and $10^6 \yr$ at a density of $100 \pcc$ (dotted lines).}
  \label{fig:cs_alt}
\end{figure*}

\subsection{Robustness of results}

The majority of our supercritical core models produce line intensities {and widths} which are incompatible with observational data. {One key} discrepancy is that these models predict CS $J=2-1$ peak intensities well above those observed, whereas subcritical models have lower gas-phase CS abundances, due to the longer duration favouring depletion onto grain surfaces, and thus lower intensities. Our chemical initial conditions are only evolved for $10^4 \yr$ at the initial core density, so it is possible that a greater degree of initial depletion could bring the models into agreement with the data. However, neither increasing the initial `settling-in' time to $10^5 \yr$, nor adopting a longer period of low-density evolution ($10^6 \yr$ at $100 \pcc$) to approximate the initial diffuse cloud conditions \citep{quenard2018}, resolves this issue. Figure \ref{fig:cs_alt} shows the resulting CS {abundances and} line profiles for the D4L1-SUP model. Changes in the initial conditions result in {factor of $\sim 3$ changes in the molecular abundance and} $\sim 0.5 \kel$ changes in the peak line intensity, nowhere near enough to bring the model into agreement with the observational data in Figure \ref{fig:csflux}. {CS is classed as a `forgetful' molecule by \citet{holdship2022}, in that its abundance rapidly equilibrates and so is relatively insensitive to the initial conditions, as are {N$_2$H$^+$, HCN and HCO$^+$}.}

{Although this kind of `settling-in' period is commonly used to generate initial conditions for models of isolated cores, it is unlikely to fully capture the previous evolution of material in a turbulent molecular cloud environment. Many prestellar cores have detectable quantities of complex molecules such as methanol and its deuterated isotopologues \citep{scibelli2020,scibelli2021,ambrose2021}, which suggests a rich chemical makeup even at this early evolutionary stage. Cores may also continue to accrete material from the parent molecular cloud over their lifetimes, rather than evolving as isolated objects \citep[e.g.][]{peretto2020,clark2021,rigby2021,anderson2021}. Properly accounting for the prior chemical evolution of core material would require a model that self-consistently forms the cores themselves from the larger structures they reside in. It is difficult to predict what impact this might have on the resulting line properties, and so we leave the topic to future work.}

Our chemical modelling involves additional assumptions beyond the initial molecular abundances, such as the cosmic ray ionization rate, isothermality, and a high degree of shielding from external radiation fields. Modifying any of these assumptions, either alone or in combination, can also have a significant impact on the resulting abundances \citep{tassis2012,priestley2018}, and so could potentially reconcile the supercritical model line {properties} with those observed. A full investigation of the sensitivity of our results to the chemical parameters is beyond the scope of this paper; we simply note that our choices are all fairly standard in the literature \citep[e.g.][]{coutens2020}. If real prestellar cores do in fact have supercritical initial mass-to-flux ratios, then some aspect of our understanding of their chemistry appears to be inaccurate.

An issue specific to using the CS $J=2-1$ line as a diagnostic is that sulphur chemistry in molecular clouds is far from completely understood \citep{holdship2019,navarro2020}. {In particular, it is common for models to invoke initial gas-phase sulphur abundances up to hundreds of times lower than the accepted solar value \citep[e.g.][]{bulut2021,navarro2021}, in order to reconcile predicted and observed line strengths of sulphur-bearing molecules, whereas we use a mostly-undepleted value (Table \ref{tab:abun}). The CS line intensities of our supercritical models could be brought into agreement with the observational data simply by reducing the initial abundance of sulphur. These reduced values require an abundant solid-phase carrier for the excess sulphur, but there is no evidence for significant sulphur depletion in either refractory \citep{jenkins2009} or icy \citep{boogert2015} material. \citet{hily2022} have recently presented evidence that the majority of sulphur exists as free atoms in the gas-phase in young prestellar cores, as we assume in our chemical modelling. Our subcritical models are in good agreement with the observed CS line strengths without invoking any additional depletion, suggesting that the `sulphur problem' may be caused by assumptions made about the physical, rather than chemical, properties of cores.}

\subsection{Comparison with prior work}

{The results presented in Figures \ref{fig:csflux} and \ref{fig:cswidth} suggest that most observed prestellar cores originate from magnetically subcritical initial conditions. This is entirely consistent with the same objects having supercritical present-day mass-to-flux ratios \citep{crutcher2012,liu2022,pattle2022}, if the observed cores represent the collapsing central regions of larger, magnetically supported structures \citep{fiedler1993}. Some estimates of core ages using chemical clocks \citep[e.g.][]{brunken2014,lin2020} return values $\gtrsim 1 \myr$, much greater than the typical free-fall times, and thus also suggestive of a subcritical mode of star formation. Other studies find ages closer to a few $10^5 \yr$ \citep{pagani2007,pagani2009,pagani2013,hily2020,caselli2022}, but these cannot be considered conclusive evidence in either direction: our D4L3-SUB model, for example, has a similar lifetime ($0.4$ versus $0.3 \myr$; Table \ref{tab:mhdprop}) to the corresponding supercritical model core, and a low inferred age could simply indicate a subcritical core seen at an early evolutionary stage. The ages obtained from molecular line observations are also highly sensitive to the details of the model employed, such as whether the density profile is taken to be static or allowed to evolve \citep{sipila2018}. These studies are all plausibly consistent with cores having initially subcritical mass-to-flux ratios.}

{Our supercritical models struggle to reproduce the low ($\lesssim 0.4 \kms$) line FWHMs typically observed in cores (Figure \ref{fig:cswidth}). Previous studies, using a variety of purely-hydrodynamical models, have not found this to be a problem \citep{rawlings1992,keto2010,keto2015}. These models typically start from much more diffuse initial conditions - the preferred model of \citet{keto2015} for L1544 is a Bonnor-Ebert sphere with a mean density $\nh \sim 100 \pcc$, resulting in lower infall velocities and correspondingly narrower lines. This configuration is in quasi-equilibrium; the inner regions collapse, while the outer parts of the core initially remain static. This is qualitatively similar to the evolution of a magnetically subcritical core \citep{fiedler1993}, and agrees with our conclusions here, in that rapidly collapsing, highly non-equilibrium models are disfavoured by the molecular line data.}

\section{Conclusions}

We have combined non-ideal MHD simulations of prestellar cores, time-dependent {gas-grain} chemistry, and line radiative transfer {modelling} to produce self-consistent synthetic molecular line observations, for a variety of initial core properties, ages and viewing angles. While line properties vary significantly - even for the same underlying physical model, {if the viewing angle or epoch of observation are different} - we find that {cores with subcritical mass-to-flux ratios} typically have lower ratios of the CS $J=2-1$ line to the N$_2$H$^+$ $J=1-0$ line, as previously found by \citet{yin2021}, due to the longer duration of collapse and corresponding increase in freeze-out of CS onto grain surfaces. {Subcritical models also have much smaller line FWHMs ($\lesssim 0.4 \kms$) than supercritical ones.}

An observational sample of cores taken from the literature falls in a region of $I_{\rm peak}({\rm CS})-I_{\rm peak}({\rm N_2H^+})$ parameter space almost exclusively occupied by models with a subcritical initial mass-to-flux ratio. {Magnetically supercritical cores struggle to reproduce the observed distribution of line strengths, {and have significantly larger line widths than nearly all the observational sample of cores}.} This {contradicts} direct measurements of the magnetic field strength {in cores}, which {generally} find supercritical present-day {mass-to-flux ratios}, and suggests that the observed cores may represent the densest central regions of larger, magnetically-supported structures. {This would imply that magnetic fields have a much more important role in star formation than is often assumed.}

\section*{Acknowledgements}
{We thank the referee for making several useful suggestions in their report, which greatly improved this paper.} {We are grateful to Michael Anderson, Nicolas Peretto, and Michiel Hogerheijde for their advice regarding hyperfine structure.} FDP is supported by the Science and Technology Facilities Council.

\section*{Data Availability}
The data underlying this article will be made available on request. The line profiles used in the analysis are available at \url{www.github.com/fpriestley/cores}. {The observational data are available in Appendix \ref{sec:obs}.}

\bibliographystyle{mnras}
\bibliography{maglines}

\appendix

\section{Observational data}
\label{sec:obs}

{We compare our model cores with observations of the CS $J=2-1$ and N$_2$H$^+$ $J=1-0$ lines towards the cores from \citet{lee1999} and \citet{tafalla2002}. For the five cores in \citet{tafalla2002}, we digitise the line profiles presented in their figures 3 and 5, and use these to determine the peak intensities and FWHMs. For \citet{lee1999}, the larger number of cores and the presentation of the data make this impracticable. We instead estimate by eye the peak intensity above the continuum of the two lines, for each of the $43$ cores in their figure 1. Most of these estimates are rounded to the nearest $0.05 \kel$, as the resolution of the data presented did not allow a more accurate determination, which results in several points in Figure \ref{fig:csflux} falling on the same horizontal or vertical line. These data are presented as antenna temperatures, which we convert to beam temperatures using an efficiency of $0.17$ taken from that paper. This does not account for the forward beam efficiency \citep[e.g.][]{barvainis1994} - including this would lower the resulting beam temperatures, worsening the discrepancy with our supercritical models and thus strengthening our conclusions. The data from \citet{tafalla2002} and \citet{lee1999} are given in Tables \ref{tab:t02} and \ref{tab:l99} respectively.}

{The definition of peak N$_2$H$^+$ intensity differs between our two samples, due to the molecule's hyperfine structure - we take the peak intensity of the strongest component detected in the \citet{tafalla2002} data, whereas \citet{lee1999} present observations of the isolated component. Additionally, our models treat all hyperfine transitions as coming from the same level. We show in Appendix \ref{sec:hfs} that the predicted peak intensity from non-hyperfine models is intermediate between, and within a factor of about two of, the isolated and the overall peak intensities from models including the hyperfine structure. Compared to the significant scatter caused by varying other model parameters, we consider this an acceptable level of accuracy.}

\begin{table}
  \centering
  \caption{{CS $J=2-1$ and N$_2$H$^+$ $J=1-0$ peak intensities and FWHMs for the five prestellar cores in \citet{tafalla2002}.}}
  \begin{tabular}{ccccc}
    \hline
    & \multicolumn{2}{c}{CS $J=2-1$} & \multicolumn{2}{c}{N$_2$H$^+$ $J=2-1$} \\
    Core & $I_{\rm peak}$/K & $W$/km s$^{-1}$ & $I_{\rm peak}$/K & $W$/km s$^{-1}$ \\
    \hline
    L1498 & $1.55$ & $0.62$ & $0.95$ & $0.31$ \\
    L1495 & $1.12$ & $0.36$ & $1.75$ & $0.32$ \\
    L1400K & $1.66$ & $0.09$ & $1.48$ & $0.25$ \\
    L1517B & $1.62$ & $0.10$ & $1.34$ & $0.29$ \\
    L1544 & $2.27$ & $0.06$ & $1.71$ & $0.34$ \\
    \hline
  \end{tabular}
  \label{tab:t02}
\end{table}

\begin{table*}
  \centering
  \caption{{CS $J=2-1$ and N$_2$H$^+$ $J=1-0$ peak intensities for the $43$ prestellar cores with detections of both in \citet{lee1999}.}}
  \begin{tabular}{cccccccccccc}
    \hline
    & \multicolumn{2}{c}{$I_{\rm peak}$/K} & & \multicolumn{2}{c}{$I_{\rm peak}$/K} & & \multicolumn{2}{c}{$I_{\rm peak}$/K} & & \multicolumn{2}{c}{$I_{\rm peak}$/K} \\
    Core & CS & N$_2$H$^+$ & Core & CS & N$_2$H$^+$ & Core & CS & N$_2$H$^+$ & Core & CS & N$_2$H$^+$ \\
    \hline
    L1333 & $0.20$ & $0.15$ & B217-2 & $0.55$ & $0.20$ & L1512-1 & $0.40$ & $0.30$ & L234E-1 & $0.45$ & $0.15$ \\
    L1521F & $0.65$ & $0.30$ & L1524-4 & $0.25$ & $0.10$ & L183B & $0.40$ & $0.40$ & L429-1 & $0.40$ & $0.05$ \\
    TMC2 & $0.80$ & $0.30$ & L1507A-1 & $0.40$ & $0.15$ & L1704-1 & $0.55$ & $0.10$ & L922-1 & $0.35$ & $0.10$ \\
    L1495 & $0.30$ & $0.25$ & CB23 & $0.40$ & $0.10$ & L1544 & $0.50$ & $0.30$ & L63 & $0.30$ & $0.60$ \\
    L1400K & $0.45$ & $0.10$ & L1622A-2 & $0.50$ & $0.30$ & L183 & $0.35$ & $0.30$ & L673-7 & $0.30$ & $0.10$ \\
    B18-5 & $0.35$ & $0.10$ & L1696A & $0.90$ & $0.40$ & L1709B-2 & $0.70$ & $0.20$ & L1049-1 & $0.20$ & $0.10$ \\
    L1399-2 & $0.20$ & $0.05$ & L1517B & $0.45$ & $0.15$ & L158 & $0.20$ & $0.25$ & L462-2 & $0.30$ & $0.10$ \\
    L1400A & $0.35$ & $0.05$ & L1622A-1 & $0.55$ & $0.25$ & L492 & $0.45$ & $0.10$ & L694-2 & $0.35$ & $0.30$ \\
    TMC1 & $0.65$ & $0.25$ & L1696B & $0.40$ & $0.50$ & L1041-2 & $0.24$ & $0.04$ & L1049-2 & $0.20$ & $0.06$ \\
    L1148 & $0.25$ & $0.05$ & L1062C-2 & $0.25$ & $0.10$ & L1197 & $0.30$ & $0.15$ & L1155C-2 & $0.22$ & $0.06$ \\
    L1251A-2 & $0.20$ & $0.05$ & CB246-2 & $0.25$ & $0.10$ & L1155C-1 & $0.35$ & $0.10$ \\
    \hline
  \end{tabular}
  \label{tab:l99}
\end{table*}

\section{Resolution tests}
\label{sec:res}

\begin{figure*}
  \centering
  \includegraphics[width=\columnwidth]{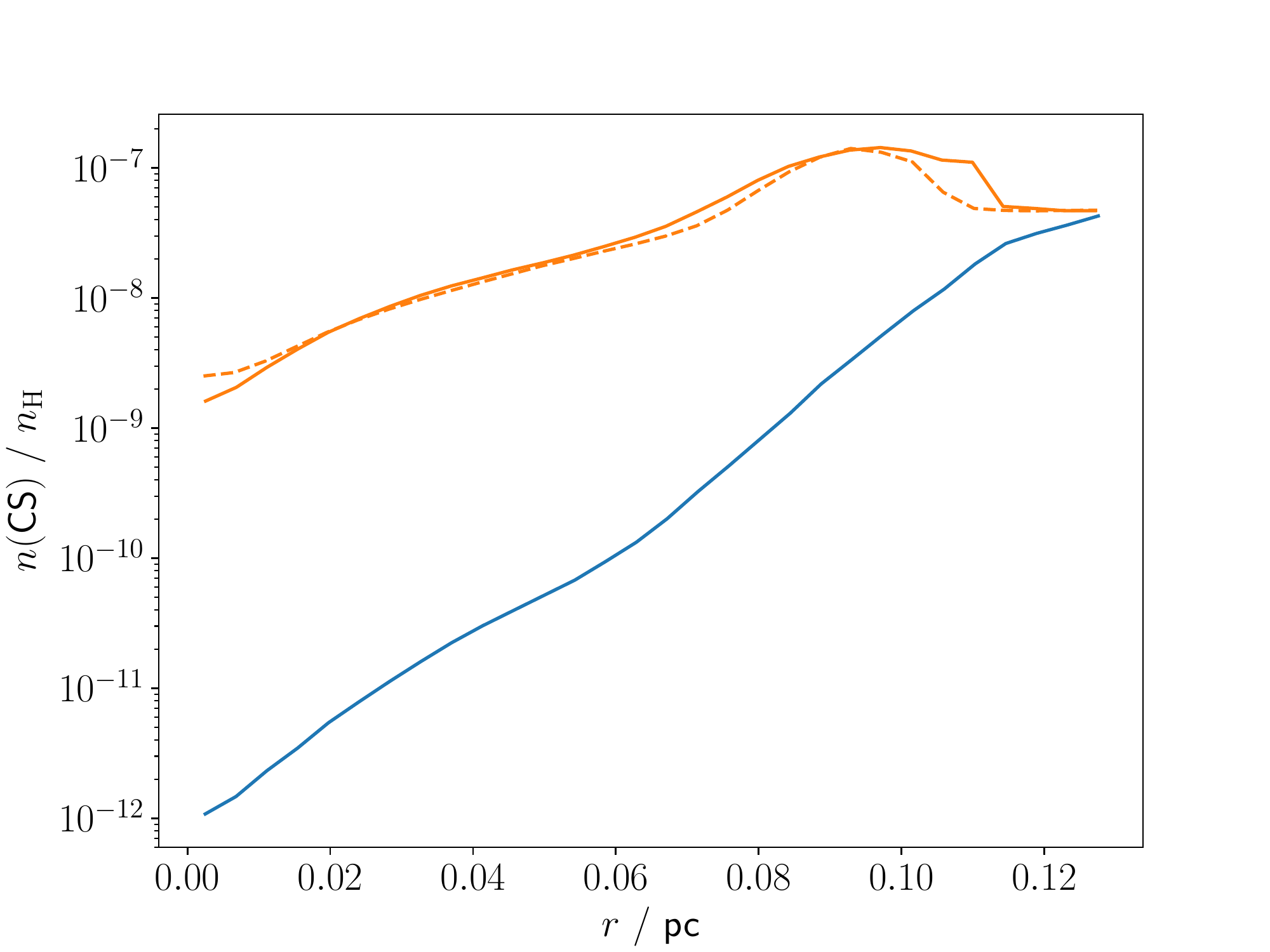}\quad
  \includegraphics[width=\columnwidth]{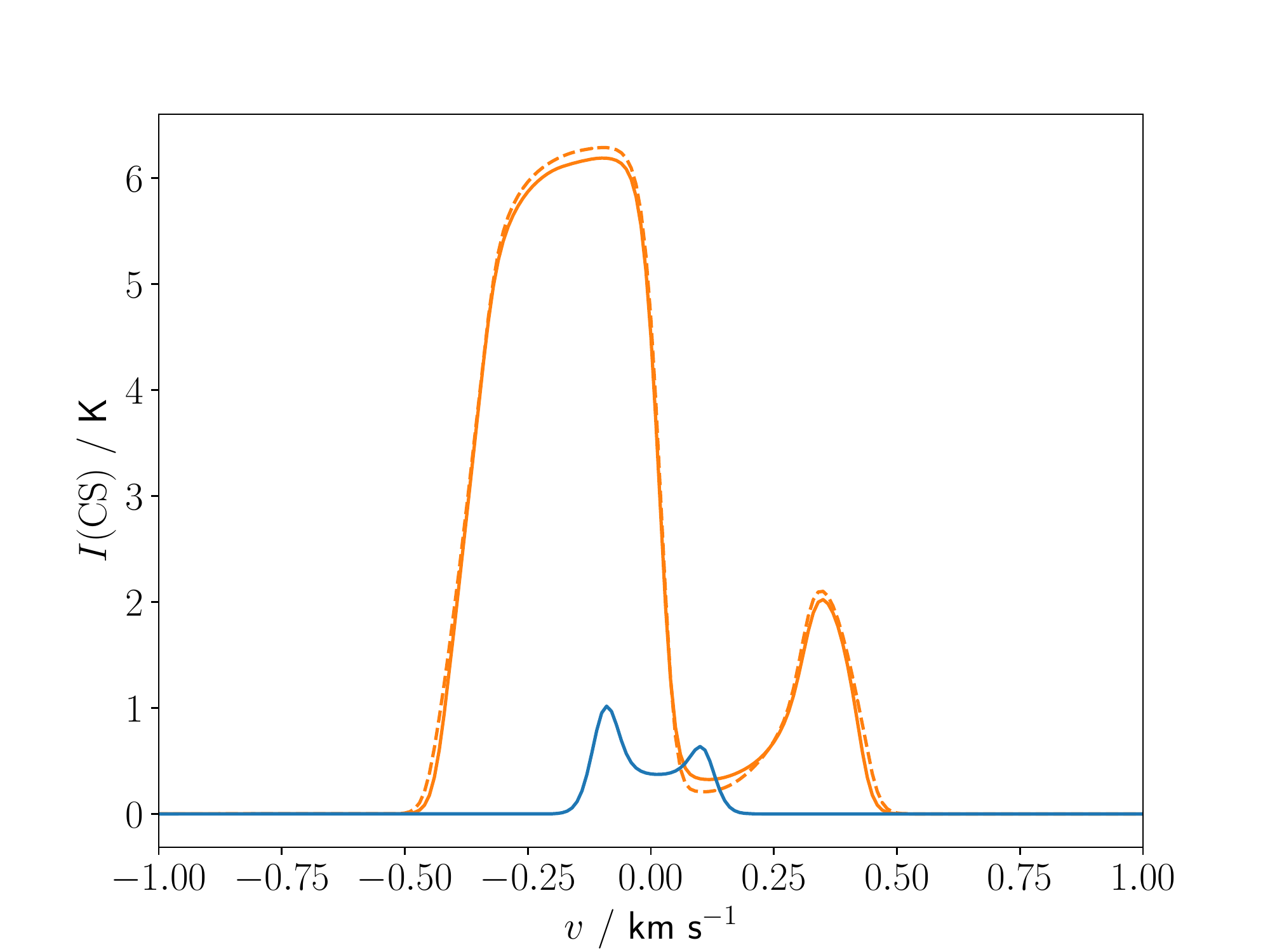}\quad
  \caption{Midplane CS abundances (left) and $J=2-1$ line intensities {viewed perpendicular to the initial magnetic field direction} (right) for the D4L1-SUB model {at $0.97 \myr$} (blue solid lines), and the D4L1-SUP model {at $0.30 \myr$} with standard (orange solid lines) and enhanced (orange dashed lines) resolution.}
  \label{fig:res}
\end{figure*}

{To investigate whether our models have sufficient resolution to produce reliable line {properties}, we repeat all stages of modelling for the D4L1-SUP model with the resolution increased by a factor of five (i.e. $10^6$ SPH particles, of which $50\,000$ are used in the chemical modelling, and $50\,000$ {\sc lime} sampling points). While the high-resolution MHD model shows some minor differences, these have a negligble impact on the quantities of interest. This is consistent with \citet{wurster2022}, who find that significant deviations due to resolution occur well beyond the densities analysed here. Figure \ref{fig:res} shows the CS midplane abundance profiles and $J=2-1$ line intensities for the normal and higher resolution models, and the (normal) subcritical model for comparison. There are small changes to both the CS abundance and line profile when the resolution is increased, but these are dwarfed by the difference between sub- and supercritical models, which are primarily driven by the different collapse timescales of the models. While our models may not be fully converged numerically, they are clearly sufficiently close to convergence for our purposes.}

\section{Hyperfine structure}
\label{sec:hfs}

\begin{figure*}
  \centering
  \includegraphics[width=\columnwidth]{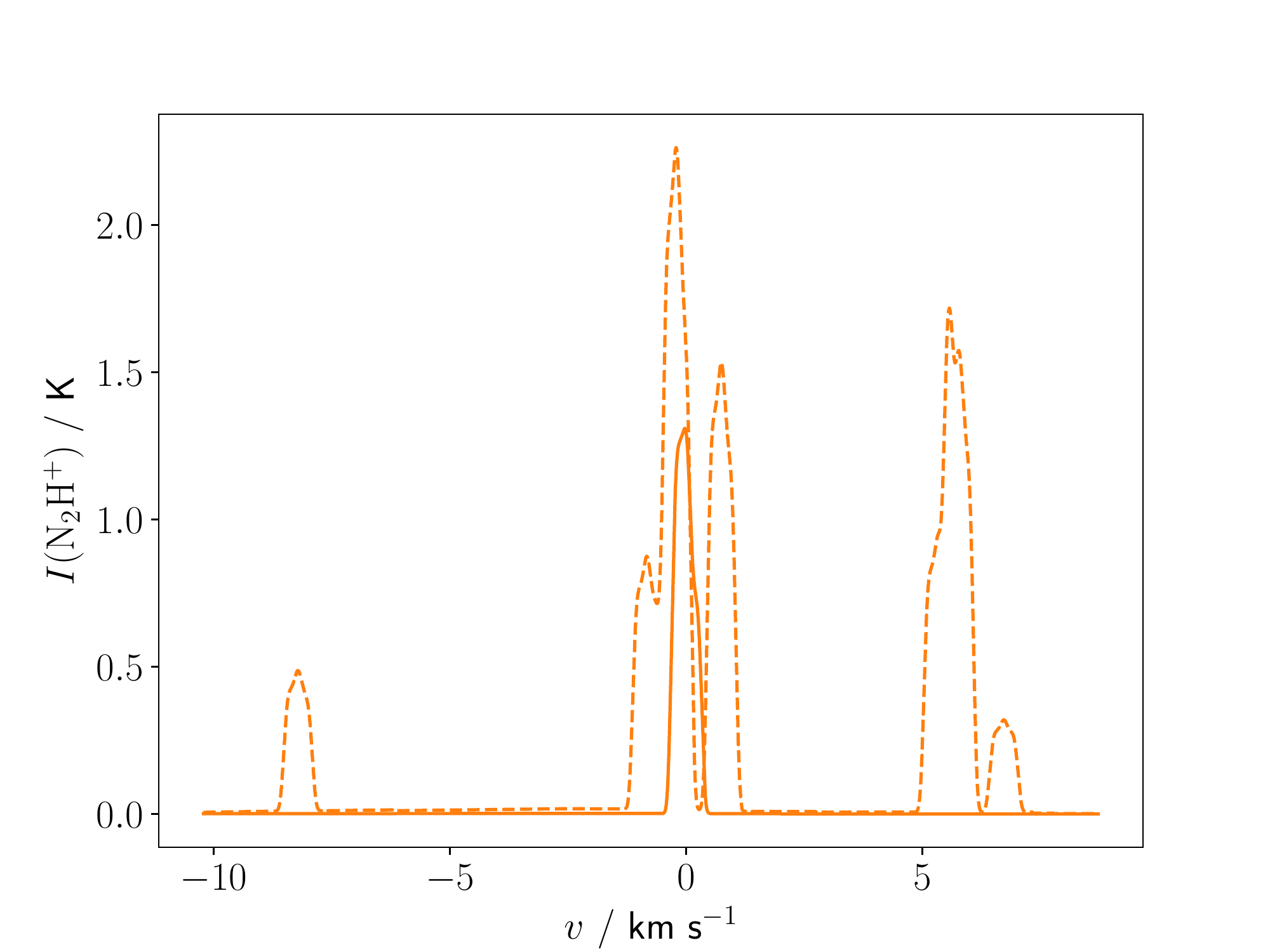}\quad
  \includegraphics[width=\columnwidth]{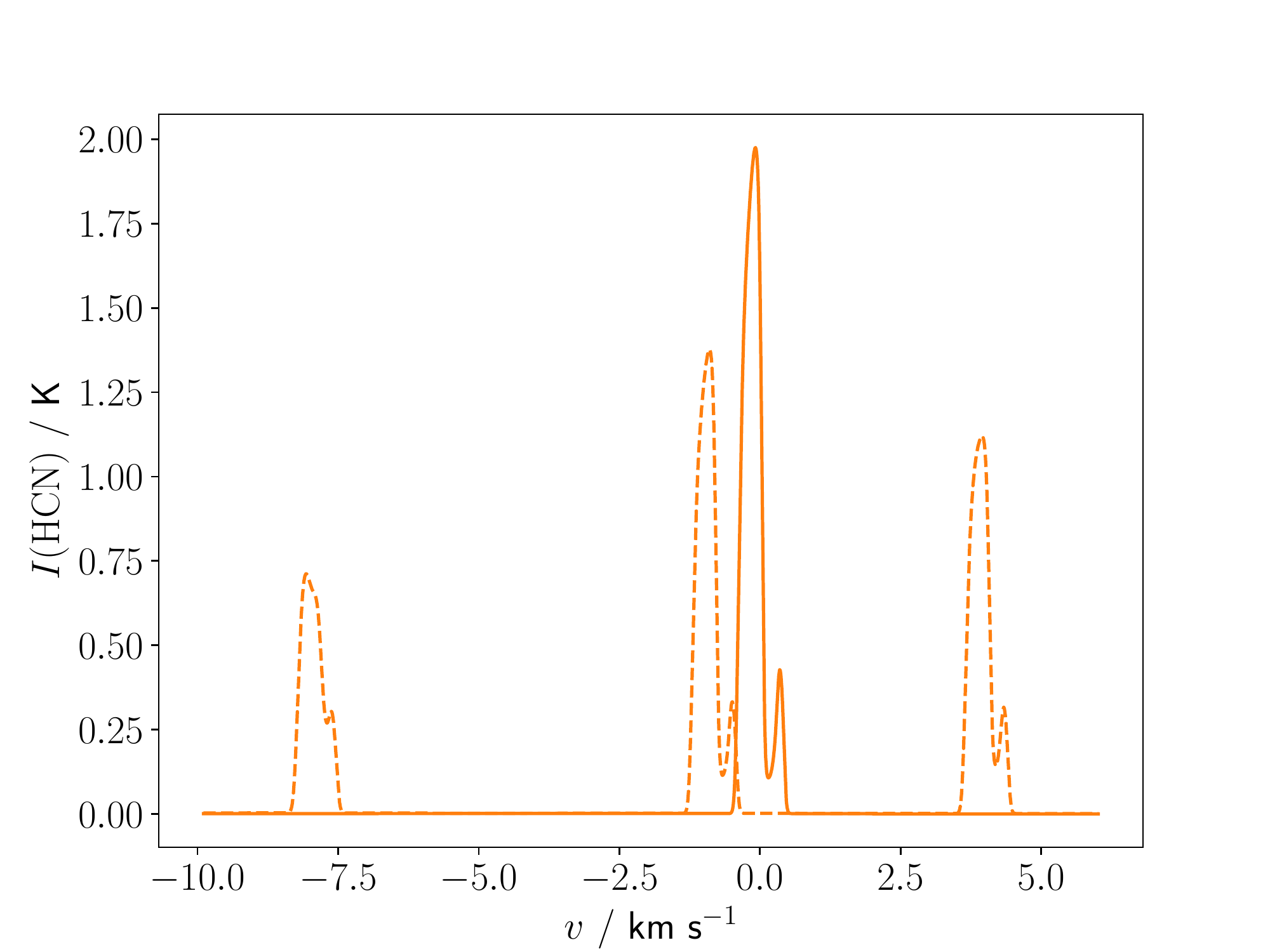}\quad
  \caption{N$_2$H$^+$ $J=1-0$ (left) and HCN $J=1-0$ (right) line profiles {viewed perpendicular to the initial magnetic field direction} for the D4L1-SUP model {at $0.30 \myr$}, with (dashed lines) and without (solid lines) hyperfine structure.}
  \label{fig:hfs}
\end{figure*}

{Figure \ref{fig:hfs} shows N$_2$H$^+$ and HCN line profiles from the D4L1-SUP model, with and without the consideration of the hyperfine structure of these transitions. There are clearly significant differences in both the line profiles and the integrated intensities, but the peak hyperfine intensities are within a factor of $\sim 2$ of the non-hyperfine values in both cases. {Different core properties, or different epochs or viewing angles of observation, can cause much larger changes in the peak intensity.} We therefore assume that the peak intensities of non-hyperfine models are sufficiently accurate for our purposes.} {The difference in line widths caused by introducing hyperfine structure is even smaller than that in intensity, and can safely be neglected.}

\bsp	
\label{lastpage}
\end{document}